\documentclass[showpacs,preprintnumbers,aps,10pt,amsmath,amssymb,prb]{revtex4}

\usepackage{graphicx}
\usepackage{subfigure}
\usepackage{dcolumn}
\usepackage{bm}

\newcommand{\kk}[0]{{\bf k}}
\newcommand{\q}[0]{{\bf q}}
\newcommand{\0}[0]{{\bf 0}}

\newcommand{\kko}[0]{{{\bf k}+{\bf k}_0}}
\newcommand{\ko}[0]{{{\bf k}_0}}
\newcommand{\Bb}[0]{{\mathcal{B}}}
\newcommand{\Ee}[0]{{\mathcal{E}}}

\newcommand{\qd}[0]{{\tilde q}}

\begin{document}

\title{Long-wavelength magnons breakdown in cubic antiferromagnets with dipolar forces at small temperature}

\author{L.\ A.\ Batalov$^1$}
\email{zlokor88@gmail.com}
\author{A.\ V.\ Syromyatnikov$^{1,2}$}
\email{asyromyatnikov@yandex.ru}
\affiliation{$^1$National Research Center "Kurchatov Institute" B.P.\ Konstantinov Petersburg Nuclear Physics Institute, Gatchina 188300, Russia}
\affiliation{$^2$Department of Physics, Saint Petersburg State University, Ulianovskaya 1, St.\ Petersburg 198504, Russia}

\date{\today}

\begin{abstract}

Using $1/S$ expansion, we discuss the magnon spectrum of Heisenberg antiferromagnet (AF) on a simple cubic lattice with small dipolar interaction at small temperature $T\ll T_N$, where $T_N$ is the Neel temperature. Similar to 3D and 2D ferromagnets, quantum and thermal fluctuations renormalize greatly the bare gapless spectrum leading to a gap $\Delta\sim \omega_0$, where $\omega_0$ is the characteristic dipolar energy. This gap is accompanied by anisotropic corrections to the free energy which make the cube edges easy directions for the staggered magnetization (dipolar anisotropy). In accordance with previous results, we find that dipolar forces split the magnon spectrum into two branches. This splitting makes possible two types of processes which lead to a considerable enhance of the damping compared to the Heisenberg AF: a magnon decay into two other magnons and a confluence of two magnons. It is found that magnons are well defined quasiparticles in quantum AF. We demonstrate however that a small fraction of long-wavelength magnons can be overdamped in AFs with $S\gg1$ and in quantum AFs with a single-ion anisotropy competing with the dipolar anisotropy. Particular materials are pointed out which can be suitable for experimental observation of this long-wavelength magnons breakdown that contradicts expectation of the quasiparticle concept.

\end{abstract}


\maketitle

\section{Introduction}

The concept of elementary excitations (quasiparticles) is a powerful approach in the modern theory of many-body systems. \cite{agd,stphys} According to this concept, each weakly excited state of a system can be represented as a set of weakly interacting quasiparticles carrying quanta of momentum $\kk$ and energy $\epsilon_\kk$. Processes of spontaneous decay of quasiparticles and interaction between them lead to a finiteness of their lifetime that is related to the quasiparticle damping $\Gamma_\kk$. It is reasonable to introduce the idea of quasiparticle only if its lifetime is sufficiently large or if the damping is much smaller than the energy ($\Gamma_\kk \ll \epsilon_\kk$). As long-wavelength quasiparticles have the smallest energies, weakly excited states of a many-body system are represented as collections of long-wavelength elementary excitations. Thus, they should be well-defined according to the quasiparticles concept. As regards short-wavelength elementary excitations, they can be defined badly or even cannot exist at all for some momenta. This situation is realized, for instance, in liquid $^4$He which has a termination point in its spectrum. \cite{pit,agd} As short-wavelength elementary excitations are normally well-defined, particular systems with overdamped short-wavelength quasiparticles have attracted much attention in recent years. \cite{1dinst,zhengprl,zhengprb,afexp,liqth1,liqth2,robin,zhitchern}

The quasiparticle concept is supported by many experiments in various systems and numerous microscopic calculations in particular models. For example, it was found in Ref.~\cite{PhysRevB.3.961} that $\Gamma_\kk \ll \epsilon_\kk$ at $k\ll1$ and $T\ll T_N$ in 3D Heisenberg antiferromagnets (AFs) with a small single-ion anisotropy, where $T_N$ is the Neel temperature. In particular, it was obtained that $\Gamma_\kk \sim \epsilon_\kk^2\tau^3\ln\tau$ at $S\sim1$ and $k\ll \tau^3$, where $\tau=T/T_N\ll1$. For large spins $S\gg1$, when the regime $T_N/S \ll T\ll T_N$ exists, the damping is estimated as $\Gamma_\kk \sim \epsilon_\kk^2\tau^2$ at $k\ll1$. 

On the other hand, it has been revealed recently \cite{syromyat2d,syromyat3d} that small long-range dipolar interaction in 2D and 3D Heisenberg ferromagnets (FMs) makes a small fraction of long-wavelength magnons to be heavily damped. 
\footnote{Notice that long-range interactions in a system are not taken into account in quite a general line of arguments supporting the quasiparticle concept. \cite{agd,stphys,hyd}}
It has been obtained that a peak appears in the ratio $\Gamma_\kk/\epsilon_\kk$ at very small but finite momentum. The peak height is of the order of unity even if the temperature is much smaller than the Curie one (i.e., when FM can be considered as a weakly excited system). This unexpected result contradicts the conventional wisdom about long-wavelength quasiparticles and expectation of the quasiparticle concept. It is demonstrated also in Refs.~\cite{syromyat2d,syromyat3d} that it is the long-range nature of the dipolar interaction that is responsible for the anomalous damping of some long-wavelength magnons. In the majority of real FM materials this effect is screened by magnetocrystalline anisotropy leading to the gap in the spectrum. Besides, it is difficult to observe the long-wavelength magnons breakdown experimentally due to very small values of the corresponding momenta. However recent progress in neutron spin-echo technique \cite{bayr,mesot} holds out hope that the corresponding measurements will be carried out in suitable FM materials. \cite{syromyat3d} 

The purpose of the present paper is to carry out similar analysis of the magnon spectrum in  Heisenberg AF with dipolar interaction on a simple cubic lattice at $T\ll T_N$ using $1/S$ expansion. We obtain that similar to 2D and 3D FMs \cite{syromyat2d,syromyat3d1} quantum and thermal fluctuations lead to anisotropic corrections to the free energy which make the cube edges easy directions for the staggered magnetization. These corrections to the free energy are naturally accompanied by appearance of a gap $\Delta$ in the magnon spectrum in the first order in $1/S$. All these phenomena are of ``order-by-disorder'' origin.

\begin{figure}
\noindent
\includegraphics[scale=0.4]{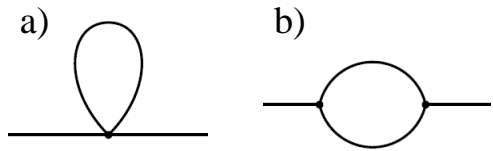}
\hfil
\caption{Diagrams of the first order in $1/S$ for self-energy parts discussed in the present paper. Diagram (a) comes from four-magnon terms \eqref{h4} in the Hamiltonian whereas (b) stems from three-magnon terms \eqref{h3}. Lines in these diagrams stand for Green's functions \eqref{gf1}.}
\label{dia}
\end{figure}

We obtain in accordance with previous results \cite{PhysRev.132.673,Harris,split} that dipolar forces split the magnon spectrum into two branches. Despite its smallness, this splitting is responsible for a considerable enhancement of the magnon damping: it opens a way for a decay of a magnon into two other spin waves and for a confluence of two magnons. These processes of decay and confluence contribute to magnon damping because the dipolar interaction leads to three-particle vertexes in the Hamiltonian. As a result, the main contribution to the damping arises in the first order in $1/S$ from the loop diagram shown in Fig.~\ref{dia}(b). This should be contrasted with Heisenberg AF, which does not have odd-particle vertexes due to the rotation invariance of the Heisenberg coupling. As a consequence, the magnon damping obtained in Ref.~\cite{PhysRevB.3.961} and mentioned above arises at finite $T$ from four-magnon vertexes in the second order in $1/S$.

Our damping calculation in the first order in $1/S$ shows that magnons are well defined quasiparticles if $S\sim1$. In particular, we obtain a peak in the ratio $\Gamma_\kk/\epsilon_\kk$ at $k\sim\Delta/D$, where $D$ is the magnon velocity, which height is proportional to $T/D\ll1$. The peak height increases considerably at $S\gg1$ in the regime $T_N/S\ll T\ll T_N$ in which case $\Gamma_\kk/\epsilon_\kk\sim{\rm const}$ and a fraction of long-wavelength magnons with $k\sim\Delta/D$ turns out to be overdamped. We demonstrate that the long-wavelength magnons breakdown arises also when a single-ion anisotropy appears in the system which competes with the anisotropy of the dipolar origin mentioned above. We argue that this phenomenon can be observed in cubic AFs $\rm TlMnF_3$ and $\rm RbMnF_3$ doped with a very small amount of cobalt. Notice that dipolar forces enhance greatly the magnon damping as compared with the purely Heisenberg AFs considered in Ref.~\cite{PhysRevB.3.961} and mentioned above. 

The rest of the present paper is organized as follows. Sec.~\ref{hamtrans} is devoted to Hamiltonian transformations and to description of the technique. Renormalization of the ground state energy and the real part of spectrum are discussed in Secs.~\ref{gsren} and \ref{realspec}, respectively. The spin-wave damping is derived in Sec.~\ref{dampsec}. The long-wavelength magnons breakdown at $S\gg1$ and in AFs with the competing single-ion anisotropy is considered in Sec.~\ref{break}. Sec.~\ref{conc} contains our conclusion. Three appendixes are added with details of calculations.

\section{Hamiltonian transformation and technique}
\label{hamtrans}

\subsection{Hamiltonian transformation}

We discuss Heisenberg AF with dipolar interaction on a simple cubic lattice which Hamiltonian has the form
\begin{equation}
\mathcal{H} = \frac{1}{2}\sum_{l\neq m}
\left(
J_{lm}\delta_{\alpha \beta} - Q_{lm}^{\alpha \beta}
\right)S_l^\alpha S_{m}^\beta,
\label{ham}
\end{equation}
where summation over repeated Greek letters is implied, $J_{lm}=J>0$ for nearest neighbors and $J_{lm}=0$ for other couples of spins,
\begin{equation}
Q_{lm}^{\alpha \beta}=\frac{\omega_0}{4\pi}\frac{3R_{lm}^\alpha R_{lm}^\beta-\delta_{\alpha \beta}R_{lm}^2}{R_{lm}^5} \label{dip_forces}
\end{equation}
is the dipolar tensor,
\begin{equation}
	\omega_0 = 4\pi \frac{(g\mu_B)^2}{v_0}
\end{equation}
is the characteristic dipolar energy that is smaller than 1 K in the majority of magnetic materials, and $v_0$ is the unit cell volume. We assume in the present paper that $\omega_0\ll J$. By taking the Fourier transformation, one has from Eq.~\eqref{ham}
\begin{equation}
\mathcal{H} = \frac{1}{2}\sum_\kk
\left(  J_\kk\delta_{\alpha \beta }-Q_\kk^{\alpha \beta}   \right)
S_\kk^\alpha S_{-\kk}^\beta , \label{ham_fourier}
\end{equation}
where $J_\kk=\sum_l J_{l m}\exp(i\kk {\bf R}_{l m})$ and $Q_\kk^{\alpha \beta}=\sum_l Q_{l m}^{\alpha \beta}\exp(i\kk {\bf R}_{l m})$. 

It is convenient to represent spin components in the local coordinate frame as follows: 
${\bf S}_l=S_l^x\hat{x}+(S_l^y\hat{y}+S_l^z\hat{z})\exp(i\kk_0{\bf R}_l)$, where $\hat{x}$, $\hat{y}$, and $\hat{z}$ are mutually orthogonal unit vectors, $\exp(i\kk_0{\bf R}_l)$ describes AF spin ordering being equal to $+1$ and $-1$ on sites belonging to different magnetic sublattices, $\ko=(\pi,\pi,\pi)$ is the AF vector, and we set the lattice spacing to be equal to unity. This representation allows one introducing only one sort of bosons via the Dyson-Maleev spin representation which has the form
\begin{eqnarray}
S_l^x &=& \sqrt{\frac{S}{2}}\left(  a_l+a_l^\dagger-\frac{a^\dagger_l a_l^2}{2S}
\right), \nonumber\\
S_l^y &=& -i\sqrt{\frac{S}{2}}\left(   a_l-a_l^\dagger-\frac{a^\dagger_l a_l^2}{2S}
\right), \label{dm}\\
S_l^z &=& S-a^\dagger_l a_l. \nonumber
\end{eqnarray}
Using the Holstein-Primakoff transformation instead of the Dyson-Maleev one \eqref{dm} does not change the results obtained below. Taking the Fourier transformation in Eqs.~\eqref{dm} and using the relation ${\bf S}_\kk = S_\kk^x\hat{x}+S_\kko^y\hat{y}+S_\kko^z\hat{z}$ one obtains from Eq.~\eqref{ham_fourier} for the Hamiltonian $\mathcal{H}=E_{gs}+\sum_{i=1}^6\mathcal{H}_i$, where 
\begin{equation}
\label{e0}
E_{gs} = -\frac12 S^2J_\0 N
\end{equation}
is the classical ground-state energy, $N$ is the number of spins in the lattice, and $\mathcal{H}_i$ denote terms containing products of $i$ operators $a^\dagger$ and $a$. In particular, $\mathcal{H}_1=0$ because it contains only $Q_{\bf 0}^{\alpha \beta}$ and $Q_\ko^{\alpha \beta}$ with $\alpha \neq \beta$ which are equal to zero. One has for the rest terms which are essential for further consideration
\begin{eqnarray}
 \label{h2}
\mathcal{H}_2 &=& \sum_{\bf k}
\left( 
E_{\bf k}a^\dagger_{\bf k}a^{}_{\bf k} + \frac{B_{\bf k}}{2}\left(a^{}_{\bf k}a^{}_{-{\bf k}}+a^\dagger_{\bf k}a^\dagger_{-{\bf k}}\right)
+\mathcal{E}_{\bf k}a^\dagger_{{\bf k}+{\bf k}_0}a^{}_{\bf k}
+ \frac{\mathcal{B}_{\bf k}}{2} a^{}_{{\bf k}}a^{}_{-{\bf k}+{\bf k}_0}
+ \frac{\mathcal{B}_{\bf k}^*}{2} a^\dagger_{{\bf k}}a^\dagger_{-{\bf k}+{\bf k}_0}
\right),\\
\mathcal{H}_3 &=& \sqrt{\frac{S}{2N}}\sum_{{\bf k}_1+{\bf k}_2+{\bf k}_3={\bf 0}}
\left( \label{h3}
iQ^{yz}_{{\bf k}_2+{\bf k}_0}a^\dagger_{{\bf k}_1} \left(a^\dagger_{{\bf k}_2} - a^{}_{-{\bf k}_2}\right)a^{}_{-{\bf k}_3}
+Q^{xz}_{{\bf k}_2}a^\dagger_{{\bf k}_1+{\bf k}_0}\left(a^\dagger_{{\bf k}_2} + a^{}_{-{\bf k}_2}\right)a^{}_{-{\bf k}_3}
\right),\\
\mathcal{H}_4 &=& \frac{1}{4N}\sum_{{\bf k}_1+{\bf k}_2+{\bf k}_3+{\bf k}_4={\bf 0}}
\left( \label{h4}
a_{{\bf k}_1}^\dagger a^{}_{-{\bf k}_2} a^{}_{-{\bf k}_3} a^{}_{-{\bf k}_4} \left(-2 J_{{\bf k}_2}+Q_{{\bf k}_2}^{xx}-Q_{{\bf k}_2+{\bf k}_0}^{yy}\right)\right.\nonumber\\
&&{}+ a_{{\bf k}_1}^\dagger a_{{\bf k}_2}^\dagger a^{}_{-{\bf k}_3} a^{}_{-{\bf k}_4} \left(-2 J_{{\bf k}_1+{\bf k}_3}+Q_{{\bf k}_1}^{xx}+Q_{{\bf k}_1+{\bf k}_0}^{yy}-2 Q_{{\bf k}_1+{\bf k}_3+{\bf k}_0}^{zz}\right)\\\nonumber
&&\left.{}-i a_{{\bf k}_1 + {\bf k}_0}^\dagger 
\left(\left(Q_{{\bf k}_2}^{xy}-Q_{{\bf k}_2+{\bf k}_0}^{xy}\right) a_{{\bf k}_2}^\dagger + a^{}_{-{\bf k}_2} \left(Q_{{\bf k}_2}^{xy}+Q_{{\bf k}_2+{\bf k}_0}^{xy}\right) \right)a^{}_{-{\bf k}_3} a^{}_{-{\bf k}_4} 
\right),
\end{eqnarray}
where
\begin{eqnarray}
E_\kk &=& SJ_{\bf 0} - \frac{S}{2}\left(Q^{xx}_\kk+Q^{yy}_\kko\right),\nonumber\\
\label{h2coef}
B_\kk &=& SJ_{\bf k} - \frac{S}{2}\left(Q^{xx}_\kk-Q^{yy}_\kko\right),\\
\mathcal{E}_\kk &=& i\frac{S}{2}\left( Q_\kko^{xy}-Q^{xy}_\kk\right),\nonumber\\
\mathcal{B}_\kk &=& i\frac S2 \left( Q_\kko^{xy}+Q^{xy}_\kk\right).\nonumber
\label{h2_denote_e}
\end{eqnarray}

\subsection{Green's functions}

It is convenient for further calculations to introduce the following set of retarded Green's functions:
\begin{align}
\label{gf1}
G(\omega,\kk)&=\langle a_\kk,a^\dagger_\kk\rangle_\omega, 
& \overline{G}(\omega,\kk)&=\langle a^\dagger_{-\kk},a_{-\kk}\rangle_\omega=G^*(-\omega,-\kk),
\nonumber\\
F(\omega,\kk)&=\langle a_\kk,a_{-\kk}\rangle_\omega,
& F^\dagger(\omega,\kk)&=\langle a^\dagger_{-\kk},a^\dagger_{\kk}\rangle_\omega=F^*(-\omega,-\kk),
\\
\mathcal{G}(\omega,\kk)&=\langle a_\kko,a^\dagger_\kk\rangle_\omega,
&\overline{\mathcal{G}}(\omega,\kk)&=\langle a^\dagger_{-\kko},a_{-\kk}\rangle_\omega=\mathcal{G}^*(-\omega,-\kk),
\nonumber\\
\mathcal{F}(\omega,\kk)&=\langle a_\kko,a_{-\kk}\rangle_\omega,
&\mathcal{F}^\dagger(\omega,\kk)&=\langle a^\dagger_{-\kko},a^\dagger_{\kk}\rangle_\omega=\mathcal{F}^*(-\omega,-\kk).\nonumber
\end{align}
We have two sets of Dyson equations for them one of which has the following form:
\begin{equation}
\label{Dyson}
\begin{pmatrix}
\overline{\Sigma}_\kk-\omega+E_\kk & B_\kk+\Pi_\kk & -\mathcal{E}_\kk+\overline{\mathcal{S}}_\kk & -\mathcal{B}_\kk+\mathcal{P}_\kk \\
B_\kk+\Pi^\dagger_\kk & \Sigma_\kk+\omega+E_\kk & \mathcal{B}_\kk+\mathcal{P}^\dagger_\kk & \mathcal{E}_\kk+\mathcal{S}_\kk \\
\mathcal{E}_\kk+\overline{\mathcal{S}}_\kko  & -\mathcal{B}_\kk+\mathcal{P}_\kko  & \overline{\Sigma}_\kko-\omega+E_\kko & B_\kko+\Pi_\kko  \\
\mathcal{B}_\kk+\mathcal{P}_\kko^\dagger & -\mathcal{E}_\kk+\mathcal{S}_\kko & B_\kko+\Pi^\dagger_\kko & \Sigma_\kko+\omega+E_\kko
\end{pmatrix}
\begin{pmatrix}
G\\F^\dagger\\\mathcal{G}\\\mathcal{F}^\dagger
\end{pmatrix}
=
\begin{pmatrix}
-1\\0\\0\\0
\end{pmatrix},
\end{equation}
where couples of self-energy parts are introduced $\Sigma_\kk=\Sigma(\omega,\kk)$ and $\overline{\Sigma}_\kk=\overline{\Sigma}(\omega,\kk)$, $\Pi_\kk=\Pi(\omega,\kk)$ and $\Pi^\dagger_\kk=\Pi^\dagger(\omega,\kk)$, $\mathcal{S}_\kk=\mathcal{S}(\omega,\kk)$ and $\overline{\mathcal{S}}_\kk=\overline{\mathcal{S}}(\omega,\kk)$, $\mathcal{P}_\kk=\mathcal{P}(\omega,\kk)$ and $\mathcal{P}^\dagger_\kk=\mathcal{P}^\dagger(\omega,\kk)$, and we use relations $\Bb^*_\kk=-\Bb_\kk=-\Bb_\kko$ and $\Ee^*_\kk=-\Ee_\kk=\Ee_\kko$ following from Eqs.~\eqref{h2coef}. 

The general solution of Eq.~\eqref{Dyson} is quite cumbersome and we do not present it here. Green's functions derived from Eq.~\eqref{Dyson} in the spin-wave approximation (i.e., at zero self-energy parts) are presented in Appendix~\ref{ap_spectr} (see Eqs.~\eqref{gfsw}). Their denominator has the form
\begin{equation}
\label{d0}
	{\cal D}^{(0)}(\omega,\kk) = \left(\omega^2-\left(\epsilon_{0\kk}^+\right)^2\right)\left(\omega^2-\left(\epsilon_{0\kk}^-\right)^2\right),
\end{equation}
where
\begin{eqnarray}
\label{specsw}
\left(\epsilon_{0\kk}^\pm\right)^2 &=& 
\frac12 \left(E_\kk^2+E_\kko^2-B_\kk^2-B_\kko^2+2\Bb_\kk^2-2\Ee_\kk^2\right)\pm \sqrt{d_\kk},\\\nonumber
\label{dk}
d_\kk &=& \frac14 \left(E_\kk^2+E_\kko^2-B_\kk^2-B_\kko^2+2\Bb_\kk^2-2\Ee_\kk^2\right)^2\\
&&{}-\left( (E_\kk+B_\kk)(E_\kko-B_\kko) +(\Ee_\kk-\Bb_\kk)^2 \right) \left( (E_\kk-B_\kk)(E_\kko+B_\kko) +(\Ee_\kk+\Bb_\kk)^2 \right)
\end{eqnarray}
and $\epsilon_{0\kk}^\pm$ give energies of two magnon branches in the spin-wave approximation (i.e., the classical magnon spectrum). It is seen that dipolar forces split the spectrum into two branches as it was pointed out before. \cite{Harris,split}  It can be shown using Eqs.~\eqref{h2coef}, \eqref{specsw}, and \eqref{dk} that $\epsilon_{0\kk}^\pm$ are invariant under replacement of $\kk$ by $\kk+\kk_0$.

To find magnon spectrum in the first order in $1/S$ (that is denoted below as $\epsilon_{1\kk}^\pm$), one has to use Green's functions \eqref{gfsw} for diagrams calculation and to consider the first $1/S$ corrections to the Green's functions denominator that has the form up to a factor
\begin{equation}
\label{d1}
	{\cal D}^{(1)}(\omega,\kk) = \left(\omega^2-\left(\epsilon_{0\kk}^+\right)^2\right)\left(\omega^2-\left(\epsilon_{0\kk}^-\right)^2\right)
	+ \Omega(\omega,\kk),
\end{equation}
where $\Omega(\omega,\kk)$ is a function linear in self-energy parts. The explicit expression for $\Omega(\omega,\kk)$ is given in Appendix~\ref{ap_spectr} (see Eqs.~\eqref{omega} and \eqref{domega}). It is also shown in Appendix~\ref{ap_spectr} that $\Omega(\omega,\kk)$ is invariant under replacement of $\kk$ by $\kk+\kk_0$. As a result $\epsilon_{1\kk}^\pm$ (like $\epsilon_{0\kk}^\pm$) have the same form in the vicinity of $\kk={\bf 0}$ and $\kk=\kk_0$. That is why we discuss below the spectrum only in the neighborhood of the point $\kk={\bf 0}$ (i.e., at $k\ll1$) bearing in mind that it has the same form near $\kk={\bf k}_0$. 

\subsection{Magnon spectrum}

Classical magnon spectrum given by Eq.~\eqref{specsw} becomes simpler in the limiting case of $k\ll1$. As it is seen from Eqs.~\eqref{h2coef}, one has to use properties of dipolar tensor at $k\ll1$ and $\kk\sim\kk_0$ to derive it. The dipolar tensor has the well-known form near the point $\kk=\0$
\begin{equation}
\label{q0}
Q_\kk^{\alpha \beta}=\omega_0\left( \frac{\delta_{\alpha \beta}}{3}-
\frac{k_\alpha k_\beta}{k^2} \right),\quad k\ll 1.
\end{equation}
We obtain numerically using the dipolar sums computation technique (see, e.g., Ref.~\cite{cohen} and references therein) at $\kk\sim\kk_0$
\begin{eqnarray}
\label{qk0}
Q^{\alpha \beta}_\kko &=&
\omega_0 \left( 
c_{xx}\left(3k_\alpha^2-k^2\right)\delta_{\alpha \beta}
+
c_{xy}k_\alpha k_\beta \left(1-\delta_{\alpha\beta}\right) 
 \right),
\quad k\ll 1,\nonumber\\
c_{xx} &\approx& 0.051,\\
c_{xy} &\approx& -0.055.\nonumber
\end{eqnarray}
In particular, it is seen from Eq.~\eqref{qk0} that $Q^{\alpha \beta}_{\kk_0}$=0. It can be shown that higher order terms in powers of $k$ in Eqs.~\eqref{q0} and \eqref{qk0} do not contribute to the results obtained in the present paper in the considered orders in $\omega_0/J$ and $k$. 

\begin{figure}
\noindent
\includegraphics[scale=0.25]{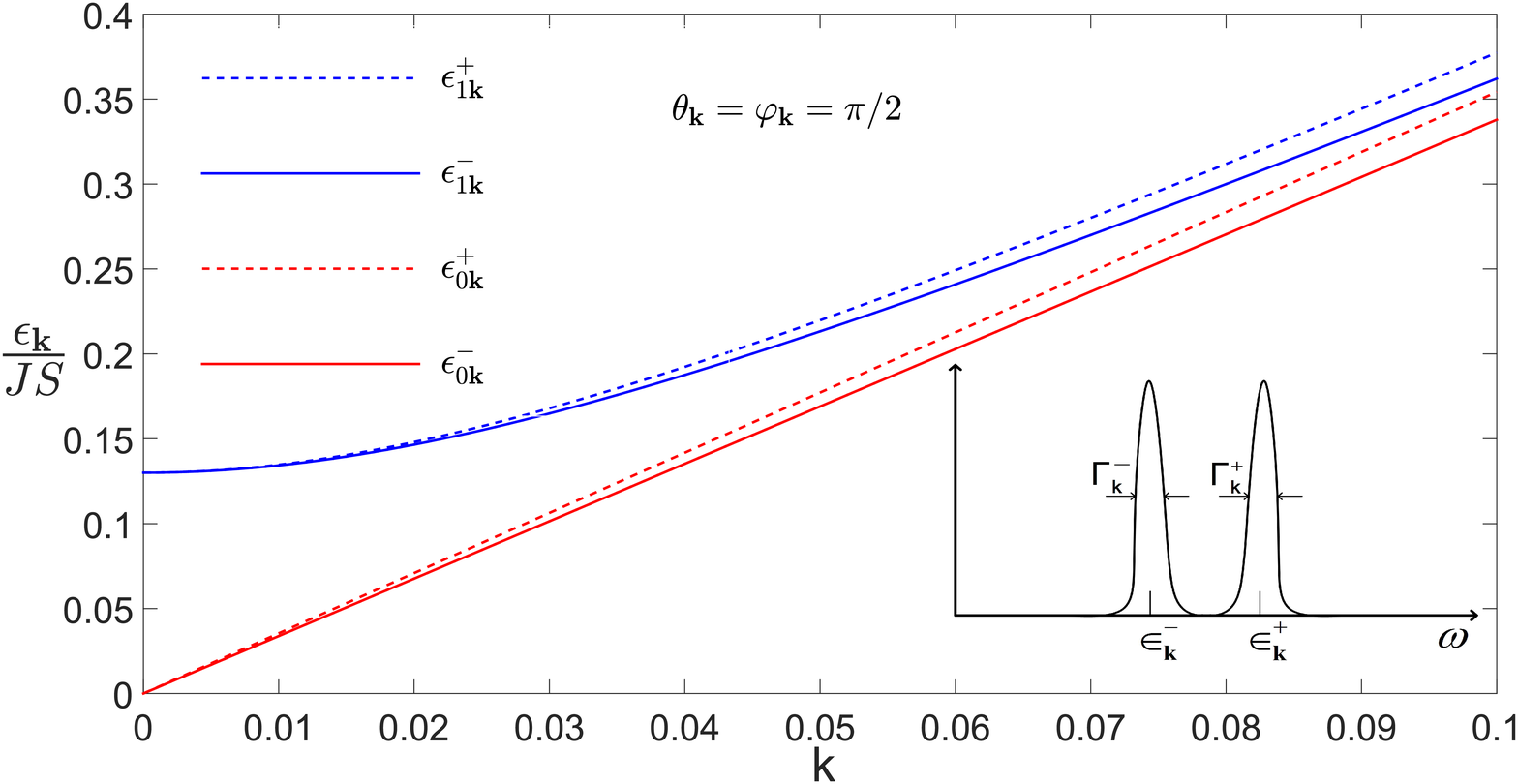}
\hfil
\caption{(Color online.) The splitting of the magnon spectrum into two branches obtained in the spin-wave approximation ($\epsilon_{0\kk}^\pm$ given by Eq.~\eqref{spectr}) and in the first order in $1/S$ ($\epsilon_{1\kk}^\pm$ given by Eq.~\eqref{specren}). Curves are drawn for $S=1/2$, $J=1$, and $\omega_0=0.4$. Inset shows a sketch of the sum of two  dynamical structure factors ${\cal S}^{xx}(\omega,\kk)+{\cal S}^{yy}(\omega,\kk)$.}
\label{sp}
\end{figure}

One obtains from Eqs.~\eqref{h2coef}, \eqref{specsw}, \eqref{dk}, \eqref{q0}, and \eqref{qk0} for the classical spectrum at $k\ll1$ in the leading order in $\omega_0/J$
\begin{eqnarray}
\label{spectr}
\epsilon_{0\kk}^\pm &=& 
D k \sqrt{1-2k^2L_2(\theta_\kk,\varphi_\kk)\pm \frac{\omega_0}{4J_{\bf 0}} L_1(\varphi_\kk)\sin^2 \theta_\kk}\approx
D k \left(1-k^2L_2(\theta_\kk,\varphi_\kk)\pm \frac{\omega_0}{8J_{\bf 0}} L_1(\varphi_\kk)\sin^2 \theta_\kk\right),\\
\label{l1}
&&L_1(\varphi_\kk) = 
\sqrt{(1+12 c_{xy})^2 + 24 (3 c_{xx}-c_{xy}) (1+18 c_{xx}+6 c_{xy}) \cos^2 2\varphi_\kk}
\approx \sqrt{4.063 + 3.95\cos 4\varphi_\kk},\\
\label{l2}
&&L_2(\theta_\kk,\varphi_\kk) = 
\frac{1}{12}\left(\cos^4\theta_\kk+\cos^2\theta_\kk\sin^2\theta_\kk+\frac18\sin^4\theta_\kk(7+\cos 4\varphi_\kk)\right) \approx L_2=0.069,
\end{eqnarray}
where $D=S\sqrt{2JJ_{\bf 0}}=SJ\sqrt{12}$ is the magnon velocity, angles $\theta_\kk$ and $\varphi_\kk$ are taken in the spherical coordinate system with $z$-axis directed along the staggered magnetization and constants $c_{xx}$ and $c_{xy}$ are defined in Eq.~\eqref{qk0}. It should be noted that the classical spectrum \eqref{spectr} is gapless. The function $L_2(\theta_\kk,\varphi_\kk)$ is very smooth: its values lie in the interval $[\frac{1}{18},\frac{1}{12}]$. That is why $L_2(\theta_\kk,\varphi_\kk)$ can be averaged over the angles and replaced by the constant $L_2$ for simplicity (see Eq.~\eqref{l2}) as it was done in Ref.~\cite{PhysRevB.3.961}. The spectrum splitting depends on the function $L_1(\varphi_\kk)$ which has the following properties: 
\begin{equation}
\label{l1prop}
0.34 \approx 1+12c_{xy} = L_1(\pi/4) \leq L_1(\varphi_\kk) \leq L_1(0) = 1+36c_{xx} \approx 2.83	
\end{equation}
and $L_1(\varphi_\kk) \approx 2\sqrt2 |\cos2\varphi_\kk|$. Spectrum \eqref{spectr} is plotted in Fig.~\ref{sp} for a particular set of parameters.

It should be noted that Eq.~\eqref{spectr} differs from the classical spectrum obtained in Ref.~\cite{Harris}. The origin of this discrepancy is in the fact that dipolar tensor components were found in Ref.~\cite{Harris} with the precision $O(k)$, whereas some quadratic in $k$ terms contribute to $\epsilon_{0\kk}^\pm$. In our notation, these are terms taken into account in Eq.~\eqref{qk0} (quadratic in $k$ terms which are omitted in Eq.~\eqref{q0} do not contribute to Eq.~\eqref{spectr}).

Magnon spectrum can be extracted from the dynamical structure factor (DSF) that is measured in neutron scattering experiment. In the spin-wave approximation, transverse DSF is determined by a linear combination of Green's functions \eqref{gf1}. In particular, DSF ${\cal S}^{xx}(\omega,\kk)$ has the form at $\kk\sim\kk_0$
\begin{eqnarray*} 
{\cal S}^{xx}(\omega,\kk)
&\propto &
{\rm Im} (G(\omega,\kk) + \overline{G}(\omega,\kk) + F(\omega,\kk) + F^\dagger(\omega,\kk) )\\
&=& \pi \frac{SJ_{\bf 0}}{\omega}\left( \delta(\omega-\epsilon_{0\kk}^+) + \delta(\omega+\epsilon_{0\kk}^+) \right)\left(1 - (1+36c_{xx})\frac{\cos(2\varphi_\kk)}{L_1(\varphi_\kk)}\right)\nonumber\\
&+&\pi \frac{SJ_{\bf 0}}{\omega}\left( \delta(\omega-\epsilon_{0\kk}^-) + \delta(\omega+\epsilon_{0\kk}^-) \right)\left(1 + (1+36c_{xx})\frac{\cos(2\varphi_\kk)}{L_1(\varphi_\kk)}\right),
\end{eqnarray*} 
where we use Eqs.~\eqref{gfsw} for Green's functions in the spin-wave approximation. The sum ${\cal S}^{xx}(\omega,\kk)+{\cal S}^{yy}(\omega,\kk)$ has the simpler dependence on angles:
\begin{eqnarray*} 
{\cal S}^{xx}(\omega,\kk)+{\cal S}^{yy}(\omega,\kk)
&\propto &
 2\pi \frac{SJ_{\bf 0}}{\omega}\left( \delta(\omega-\epsilon_{0\kk}^+) + \delta(\omega+\epsilon_{0\kk}^+) +\delta(\omega-\epsilon_{0\kk}^-) + \delta(\omega+\epsilon_{0\kk}^-)\right),
\end{eqnarray*} 
where $\kk\sim\kk_0$. It is sketched in the inset of Fig.~\ref{sp}, where we take into account that delta-peaks are replaced by Lorentzian functions due to the magnon damping derived below.

\section{The ground state energy renormalization}
\label{gsren}

The classical ground state of the model \eqref{ham} is continuously degenerate: the staggered magnetization can have arbitrary direction as it is seen from Eq.~\eqref{e0}. It is well known that quantum fluctuations can give anisotropic corrections to the ground state energy selecting a limiting number of states (``order-by-disorder'' effect). These quantum corrections are proportional in our case to sums over momenta containing components of the dipolar tensor $Q_\kk^{\alpha \beta}$ and depend consequently on the direction of the quantized axis relative to the lattice. Thus, one should bear in mind in the subsequent calculations what is the easy direction of magnetization in the ground state. Using Eq.~\eqref{h2} for the biquadratic part of the Hamiltonian and Eqs.~\eqref{gfsw} for Green's functions, we obtain after tedious calculation the following anisotropic part of the first $1/S$ correction to the ground state energy $E_{gs}$:
\begin{eqnarray} 
\label{de}
\frac{\Delta E_{gs}}{N} &=& C\frac{S\omega_0^2}{J}
\left(\gamma_x^2\gamma_y^2+\gamma_x^2\gamma_z^2+\gamma_z^2\gamma_y^2\right),\\
\label{c}
C &=& \frac{J}{16\omega_0^2}\frac1N\sum_\q \label{c_an} \frac{\left(J_\0-J_\q\right)^2\left(\left(Q_\q^{xx}-Q_\q^{yy}\right)^2-4\left(Q_\q^{xy}\right)^2\right)}{\left(J_\0^2-J_\q^2\right)^{3/2}}
\approx 0.0022,
\end{eqnarray}
where $\gamma_i$ are direction cosines of the staggered magnetization relative to axes which are parallel to cube edges. Components of the dipolar tensor in Eq.~\eqref{c_an} are taken relative to these axes. The constant $C$ has been calculated numerically using the procedure of dipolar sums computation \cite{cohen}. This computational technique is required because momenta $q\sim1$ give the main contribution to the sum in Eq.~\eqref{c} and one cannot use Eqs.~\eqref{q0} and \eqref{qk0}. As $C>0$, cube edges are easy directions for the staggered magnetization.

It is shown in the next section that similar to FMs with dipolar interaction \cite{syromyat2d,syromyat3d1} the fluctuation-induced anisotropy \eqref{de} is naturally accompanied by the fluctuation-induced gap in the spectrum. Then, both the anisotropy and the gap have the ``order-by-disorder'' origin.

\section{Renormalization of the real part of the spectrum}
\label{realspec}

Let us discuss renormalization of the real part of the spectrum stemming from diagrams of the first order in $1/S$ shown in Fig.~\ref{dia}. Lines in these diagrams stand for bare Green's functions introduced in Eqs.~\eqref{gf1} (see Eqs.~\eqref{gfsw} for their explicit form in the spin-wave approximation). Each self-energy part arising in the Dyson equation \eqref{Dyson} receives its own contribution from the diagrams.

As can be seen from results below, it is more convenient to discuss renormalization of the real part of the spectrum square for which we have from Eq.~\eqref{d1}
\begin{equation}
\label{specsq}
\left(\epsilon_{1\kk}^\pm\right)^2
=
\left(\epsilon_{0\kk}^\pm\right)^2 \mp \frac{{\rm Re}\Omega\left(\omega=\epsilon_{0\kk}^\pm+i\delta,\kk\right)}{2\sqrt{d_\kk}},
\end{equation}
where the last term is given by Eq.~\eqref{omega1}. The Hartree-Fock diagram presented in Fig.~\ref{dia}(a) originates from four-magnon terms \eqref{h4} in the Hamiltonian. After simple calculations we obtain in the leading orders in $k$ and $\omega_0/J$ for the contribution to $\Omega(\omega,\kk)$
\begin{eqnarray}
\label{o4}
\mp\frac{\Omega^{(4)}(\epsilon_{0\kk}^\pm,\kk)}{2\sqrt{d_\kk}} &=& 
(Dk)^2\frac{1}{S N}\sum_\q \frac{J_\0-\sqrt{J_\0^2-J_\q^2}}{J_\0}\left(1+2\mathcal{N}_\q\right)\\
&&{} 
+
\frac{J_\0 S}{4N}\sum_\q \frac{(J_\0-J_\q)^2((Q_\q^{xx}-Q_\q^{yy})^2+2(Q_\q^{xy})^2)}{\left(J_\0^2-J_\q^2\right)^{3/2}}(1+2\mathcal{N}_\q)
+
\frac{J_\0 S}{2N}\sum_\q \frac{Q^{xy}_\q Q^{xy}_{\bf{q}+\ko}}{\sqrt{J_\0^2-J_\q^2}}(1+2\mathcal{N}_\q),\nonumber
\end{eqnarray}
where $\mathcal{N}_\q=(\exp(\epsilon_\q/T)-1)^{-1}$ and $\epsilon_\q=S\sqrt{J_{\bf 0}^2-J_{\bf q}^2}$. One concludes comparing Eqs.~\eqref{spectr} and \eqref{o4} that the first term in Eq. \eqref{o4} leads to the well known renormalization of the magnon velocity $D$
\begin{eqnarray}
\label{dren}
D\rightarrow 
D\left( 1 + \frac{1}{2S N}\sum_\q \frac{J_\0-\sqrt{J_\0^2-J_\q^2}}{J_\0}(1+2\mathcal{N}_\q) \right)
\approx D\left( 1+\frac{1}{2S} \left(0.097 + \frac{4\zeta(3)}{\pi^2}\left(\frac{T}{D}\right)^3  \right)\right),
\end{eqnarray}
where $\zeta(x)$ is the Riemann zeta function and we assume that $S\sim1$ (so that $T\ll D$ at $T\ll T_N$). The second and the third terms in Eq.~\eqref{o4} contribute to the spin-wave gap.

The loop diagram shown in Fig.~\ref{dia}(b) comes from three-magnon terms \eqref{h3} in the Hamiltonian. As a result of simple but tedious calculations we obtain for the contribution to the real part of $\Omega(\omega,\kk)$ in the leading orders in $k$ and $\omega_0/J$
\begin{equation}
\label{o3}
\mp\mathrm{Re}\frac{\Omega^{(3)}(\epsilon_{0\kk}^\pm,\kk)}{2\sqrt{d_\kk}}
=
-\frac{3J_\0 S}{2N}\sum_\q \frac{(J_\0-J_\q)^2(Q_\q^{xy})^2}{\left(J_\0^2-J_\q^2\right)^{3/2}}(1+2\mathcal{N}_\q)-
\frac{J_\0 S}{2N}\sum_\q \frac{Q^{xy}_\q Q^{xy}_{\bf{q}+\ko}}{\sqrt{J_\0^2-J_\q^2}}(1+2\mathcal{N}_\q),
\end{equation}
where we set $\kk=\0$ under sums because the summation over $q\sim1$ gives the main contribution. 

One obtains in the first order in $1/S$ from Eqs.~\eqref{spectr}, \eqref{specsq}, \eqref{o4} and \eqref{o3} the following expression for the spectrum at $S\sim1$ and $k\ll1$:
\begin{eqnarray}
\label{specren}
\epsilon_{1\kk}^\pm =
\sqrt{(D k)^2 \left(1-2L_2k^2 \pm \frac{\omega_0}{4J_{\bf 0}} L_1(\varphi_\kk)\sin^2 \theta_\kk\right)+\Delta^2},
\end{eqnarray} 
where we imply the small renormalization of the magnon velocity \eqref{dren},
\begin{equation}
\label{gap}
\Delta = \sqrt{24 CS}\omega_0
\end{equation}
is the gap in the spectrum, and the constant $C$ is given by Eq.~\eqref{c}. Notice that thermal corrections to the gap are negligibly small at $S\sim1$. Eq.~\eqref{specren} is plotted in Fig.~\ref{sp} for a specific set of parameters. Spectrum \eqref{specren} has the following form in the two limiting cases:
\begin{equation}
\label{specfin}
\epsilon_{1\kk}^\pm=\begin{cases}
\Delta, & k\ll \Delta/D,\\
\displaystyle D k\left( 1-L_2k^2+\frac{\Delta^2}{2(Dk)^2} \pm \frac{\omega_0}{8J_{\bf 0}} L_1(\varphi_\kk)\sin^2\theta_\kk \right), & k\gg \Delta/D.
\end{cases}
\end{equation}

As it was done in Refs.~\cite{syromyat3d1,syromyat2d} for 2D and 3D FMs with dipolar forces, it can be shown that coincidence is not accidental of the numerical constants $C$ in expressions for the anisotropic correction to the ground state energy \eqref{de} and to the gap \eqref{gap}. Namely, the anisotropy in the Hamiltonian of the type 
$C\frac{S\omega_0^2}{J}\sum_i
\left((S_i^x)^2(S_i^y)^2 + (S_i^x)^2(S_i^z)^2+(S_i^y)^2(S_i^z)^2\right)/S^4$ 
(cf.\ Eq.~\eqref{de}), where $C$ is a positive constant, leads to the gap in the classical spectrum of the form \eqref{gap} if $S\gg1$.

As is seen from Eqs.~\eqref{spectr} and \eqref{specfin}, the spectrum renormalization is very small at $k\gg\Delta/D$ whereas quantum fluctuations change it drastically at $k\ll\Delta/D$. One has to take into account this renormalization when discussing the spin-wave damping. Then, we carry out below self-consistent calculations of the damping. Notice that such self-consistent consideration leads to the same result \eqref{specren} for the real part of the spectrum.

\section{Magnon damping}
\label{dampsec}

It is well known that the magnon damping arises in Heisenberg non-frustrated AFs at $T\ne0$ in the second order in $1/S$ and there is no damping at $T=0$. \cite{PhysRevB.3.961} Dipolar forces give rise to the finite damping at $T\ge0$ in the first order in $1/S$ due to the three-magnon interaction \eqref{h3} that leads to the loop diagram shown in Fig.~\ref{dia}(b). 

\begin{figure}
\includegraphics[scale=0.4]{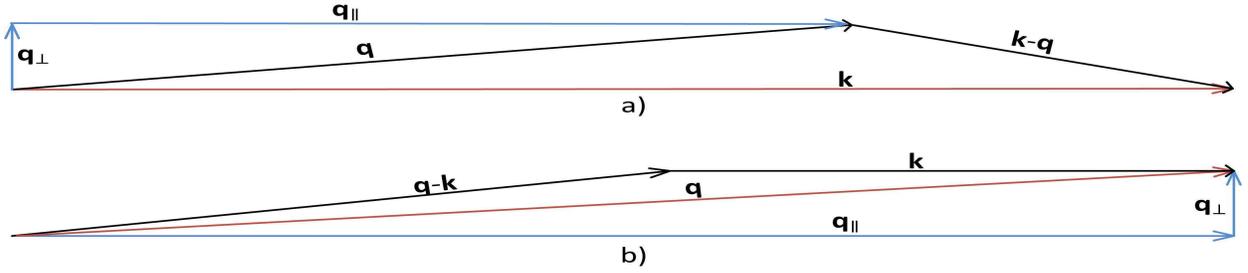}
\caption{(Color online.) Momenta $\q$, $\kk$, and $\q-\kk$ (or $\kk-\q$) of three magnons in the decay (a) and confluence (b) processes which are given by Eqs.~\eqref{decay} and \eqref{conf}, respectively. Components $\q_\|$ and $\q_\perp$ of $\q$ are also shown which are parallel and perpendicular to $\kk$, correspondingly.}
\label{vec}
\end{figure}

Contributions from the diagram presented in Fig.~\ref{dia}(b) to the imaginary part of each self-energy part contain delta-functions describing the magnon decay and the confluence of two magnons (it is clear from the explicit form of the bare Green's functions given by Eqs.~\eqref{gfsw}). For a magnon with momentum $\kk$, one has $2^3=8$ possible decay processes of the type
\begin{equation}
\label{decay}
	\epsilon_\kk^\pm-\epsilon_\q^\pm-\epsilon_{\kk-\q}^\pm=0
\end{equation}
which arise at any $T$ and 8 confluence processes
\begin{equation}
\label{conf}
	\epsilon_\kk^\pm-\epsilon_\q^\pm+\epsilon_{\kk-\q}^\pm=0
\end{equation}
which exist at $T\ne0$ only. Their contributions to the imaginary part of $\Omega(\omega,\kk)$ determining the damping are not equal and depend on $T$ and $\bf k$. Figs.~\ref{vec}(a) and \ref{vec}(b) illustrate Eqs.~\eqref{decay} and \eqref{conf}, respectively. The reader is referred to Appendix~\ref{ap_damp} for a detailed analysis of Eqs.~\eqref{decay} and \eqref{conf}.

One obtains from Eq.~\eqref{d1} for the magnon damping in the first order in $1/S$
\begin{eqnarray}
\label{dampdef}
\Gamma_{\kk}^\pm 
&=& 
\pm\frac{\mathrm{Im} \Omega(\omega=\epsilon_{1\kk}^\pm+i\delta,\kk)}{4\sqrt{d_\kk}\epsilon_{1\kk}^\pm}.
\end{eqnarray}
As it was mentioned above, one has to carry out self-consistent calculations to find $\Gamma_{\kk}^\pm $ due to the considerable renormalization of the real part of the spectrum by fluctuations at $k\lesssim\Delta$. The general expression for $\Gamma_{\kk}^\pm $ and corresponding calculations are rather cumbersome. Fortunately, the results become quite compact in limiting cases which are important for our consideration and which we discuss below.

We assume in this section that $S\sim1$. As a consequence, it is implied that $D\sim T_N$ and $\Delta\sim\omega_0$.

\subsection{$T=0$}
\label{damt=0}

As it is discussed in Appendix~\ref{ap_damp} in more detail, the decay processes contribute at $T=0$ to the damping of ``+''-magnon branch only whereas the ``$-$''-branch has infinite lifetime:
\begin{equation}
	\Gamma_{\kk}^- = 0.
\end{equation}
As a result of simple but tedious calculations we obtain 
at 
$k \gg \sin\theta_\kk\sqrt{\frac{\omega_0}{J}} $
and
$\sin\theta_\kk \gg \sqrt{\frac{\Delta}{S\omega_0}}$
\begin{eqnarray}
\label{g+t02}
\Gamma_{\kk}^+ &=& 
k\frac{\omega_0^3}{J^2} 
\frac{(1+12c_{xy})^2}{27648\pi L_2}
L_1(\varphi_\kk) \sin^4\theta_\kk \cos^2\theta_\kk
\approx
k\frac{\omega_0^3}{J^2} 
1.9\cdot10^{-5}
L_1(\varphi_\kk)\sin^4\theta_\kk \cos^2\theta_\kk,
\end{eqnarray}
and $\Gamma_{\kk}^+$ is negligible for other $k$ and $\theta_\kk$. Thus, one concludes from Eq.~\eqref{g+t02} that $\Gamma_{\kk}^+ \ll \epsilon_{\kk}^+$ at $T=0$. 

Temperature corrections to Eq.~\eqref{g+t02} become important at $T\gg S\omega_0\sin^2 \theta_\kk/k^2$. The order of magnitude of these corrections can be obtained by multiplying of Eq.~\eqref{g+t02} by
\begin{equation}
\label{tcordecay}
\frac{Tk^2}{S\omega_0\sin^2 \theta_\kk}
\ln\left( \frac{S\omega_0}{\Delta} \sin^2 \theta_\kk \right).
\end{equation}

\subsection{$T\ne0$}
\label{damtne0}

\begin{figure}
\includegraphics[scale=0.2]{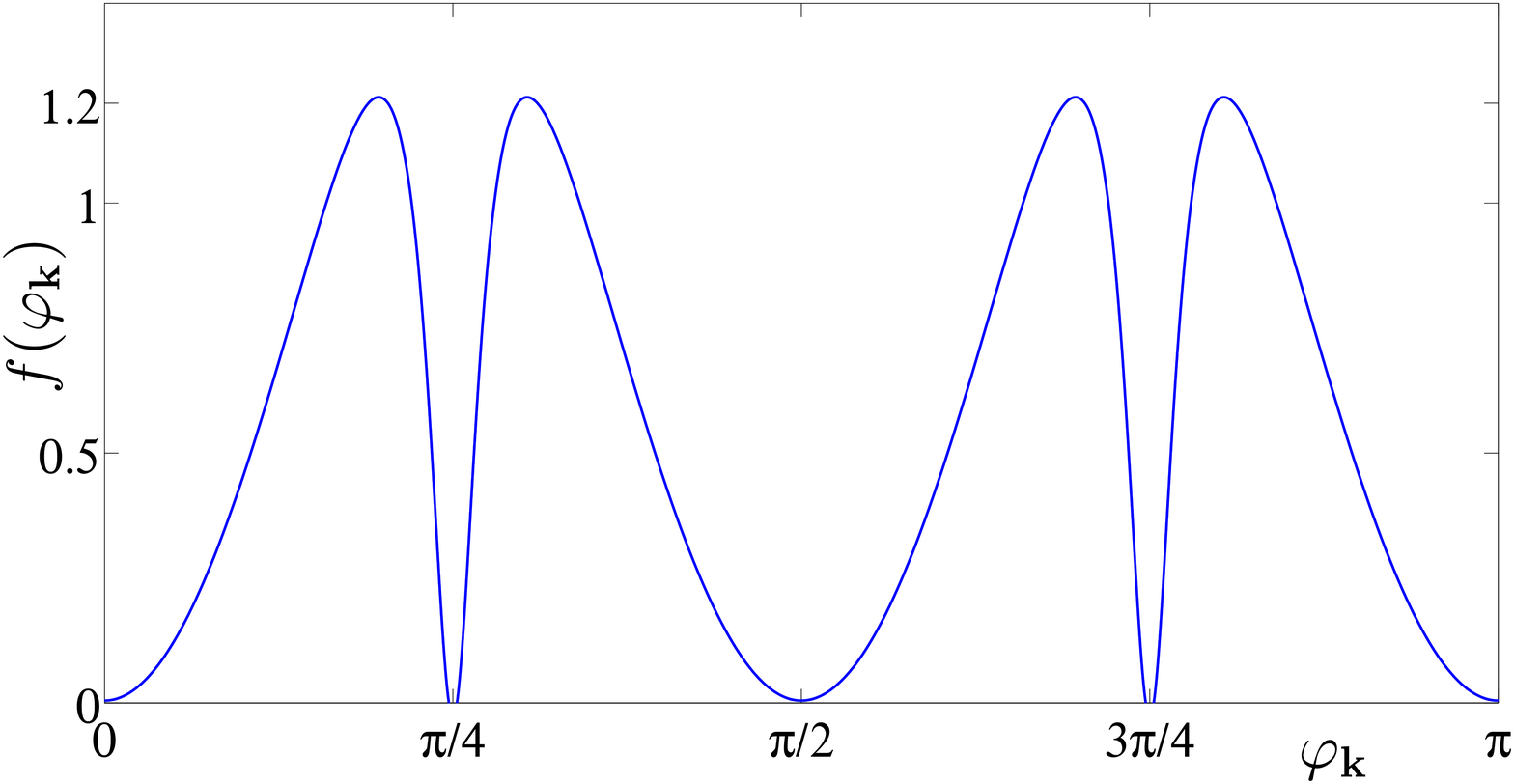}
\caption{(Color online.) Graphic of the function $f(\varphi_\kk)$ given by Eq.~\eqref{f} that determines dependence of the damping on azimuthal angle $\varphi_\kk$ (see Eqs.~\eqref{g+tn01}, \eqref{g-tn01}, \eqref{g+tn02}, \eqref{g+tn0comp}, and \eqref{g-tn0comp}).}
\label{ffig}
\end{figure}

Confluence processes give the main contribution to the damping when
$
\sin\theta_\kk\sqrt{\frac{\omega_0}{J}}
\gg
k
\gg
\frac{\Delta}{D}
$
and $\frac{S\omega_0}{k}\sin^2\theta_\kk \ll T\ll D$:
\begin{eqnarray}
\label{g+tn01}
\Gamma_{\kk}^+ &=& 
\frac{\omega_0^2}{kJ}
\frac TD
\frac{(1+12 c_{xy})^2}{2304\pi} 
A^3\sqrt{1-B}(4-B)
\left(
1-
\frac{ f(\varphi_\kk)}{4(1+18 c_{xx}+6 c_{xy})}
\right)
\sin^2 2\theta_\kk,\\
\label{g-tn01}
\Gamma_{\kk}^- &=& 
\Gamma_{\kk}^+
\frac{f(\varphi_\kk)}{4(1+18 c_{xx}+6 c_{xy})-f(\varphi_\kk)},
\end{eqnarray}
where the non-negative function
\begin{equation}
\label{f}
	f(\varphi_\kk) = \frac{(1+36 c_{xx}-L_1(\varphi_\kk)) (L_1(\varphi_\kk)-1-12 c_{xy})}{L_1(\varphi_\kk)}
\end{equation}
is introduced (cf.\ Eq.~\eqref{l1prop}) which graphic is shown in Fig.~\ref{ffig} and
\begin{eqnarray}
\label{a}
	A &=& \frac{\omega_0}{144L_2Jk} L_1(\varphi_\kk)\sin^2\theta_\kk
	\approx 0.1 \frac{\omega_0}{Jk} L_1(\varphi_\kk)\sin^2\theta_\kk, \\
\label{b}
	B &=& \frac{288 L_2 \Delta^2 }{S^2\omega_0^2L_1(\varphi_\kk)^2\sin^4\theta_\kk}
	= \frac{6912 C L_2}{SL_1(\varphi_\kk)^2\sin^4\theta_\kk}
	\approx \frac{1.05}{SL_1(\varphi_\kk)^2\sin^4\theta_\kk}.
\end{eqnarray}
Eqs.~\eqref{g+tn01} and \eqref{g-tn01} are valid for $B<1$. Momenta of summation $q\gg k$ give the main contribution to Eqs.~\eqref{g-tn01} and \eqref{g+tn01}.

For larger momenta, one obtains when
$k \gg \sin\theta_\kk\sqrt{\frac{\omega_0}{J}} $,
$\sin\theta_\kk \gg \sqrt{\frac{\Delta}{S\omega_0}}$,
and $T\gg Dk$
\begin{eqnarray}
\label{g+tn02}
\Gamma_{\kk}^+ &=&
k^2 \frac{\omega_0^2}{J}
\frac TD
\frac{(1+12c_{xy})^2}{768\pi}
\sin^2 2\theta_\kk\nonumber\\
&&\times
\left(
\ln\left( \frac{S\sqrt{J\omega_0}}{\Delta}k \sin \theta_\kk \right)
+
\left(\frac{1-12c_{xy}}{1+12c_{xy}}\right)^2
\left(
1-
\frac{ f(\varphi_\kk)}{4(1+18 c_{xx}+6 c_{xy})}
\right)
\ln\left( \frac{\sqrt B}{1-\sqrt{1-B}} \right)
\right),\\
\label{g-tn02}
\Gamma_{\kk}^- &=& 
k^2\frac{\omega_0^2}{J}
\frac TD
\frac{(1+12 c_{xy})^2}{768\pi} 
\ln\left( \frac{\sqrt B}{1-\sqrt{1-B}} \right)
\sin^2 2\theta_\kk,
\end{eqnarray}
where decay and confluence processes lead to the first and to the second terms in the last brackets in Eq.~\eqref{g+tn02}, respectively, and confluence processes determine $\Gamma_{\kk}^-$.

\begin{figure}
\noindent
\includegraphics[scale=0.5]{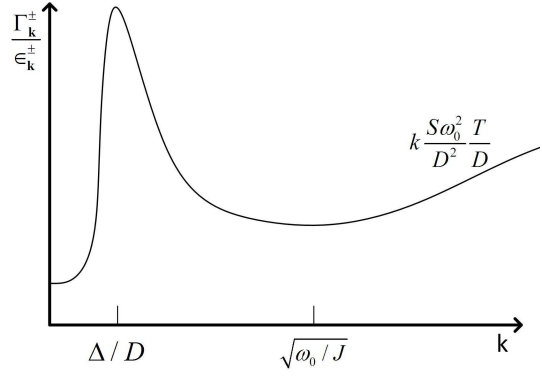}
\hfil
\caption{Sketch of the ratio $\Gamma_{\kk}^\pm/\epsilon_{\kk}^\pm$ at $T\gg S\sqrt{J\omega_0}$. Asymptotic at $k\gg \sqrt{\omega_0/J}$ is also shown. The peak hight at $k\sim \Delta/D$ is proportional to $T/D\ll1$ for quantum spins and it is given by a constant for $S\gg1$ at $D\ll T\ll T_N$. The peak height in quantum AF can reach a value of the order of unity if the gap is sufficiently reduced by the anisotropy competing with the dipolar one \eqref{de} (see discussion in Sec.~\ref{companis}).}
\label{zatuch}
\end{figure}

It is seen from Eqs.~\eqref{g+tn01}--\eqref{g-tn01} and \eqref{g+tn02}--\eqref{g-tn02} that the damping $\Gamma_{\kk}^\pm$ decreases as $k^2$ upon $k$ decreasing down to $k\sim \sqrt{\frac{\omega_0}{J}}$ in which region this decreasing changes into rising that has the form $1/k^4$. This rising takes place up to $k\sim\Delta/D$ near which point the increasing turns into a rapid fall due to the gap in the spectrum. Thus, $\Gamma_{\kk}^\pm$ has a peak at $k\sim \Delta/D$ which height is of the order of $\omega_0T/D$ and the ratio $\Gamma_{\kk}^\pm/\epsilon_\kk^\pm$ is proportional to $T/D\ll1$ at the peak position (see Fig.~\ref{zatuch}). Thus, one concludes that magnons are well defined quasiparticles in quantum AF with dipolar forces at $T\ll T_N$.

\section{Possibility of the magnon breakdown}
\label{break}

We obtain in the previous section that the damping rising upon $k$ decreasing stops at $k\sim\Delta/D$ due to the gap in the spectrum. One infers that a reduction of the gap value could keep the damping increasing and lead to the long-wavelength magnon breakdown. We discuss in this section two possibilities of the gap decreasing. First, we consider large spins $S\gg1$. As the gap \eqref{gap} obtained above is of the next order in $1/S$ as compared to the bare spectrum, the gap value can be reduced by increasing $S$. The second way to decrease the gap value is to take into account a magnetocrystalline anisotropy competing with the dipolar one \eqref{de} (i.e., the magnetocrystalline anisotropy favoring cube space diagonals rather than cube edges). In this case, a negative contribution arises to the gap \eqref{gap} which can decrease the gap considerably (see Appendix~\ref{defect} for a more detail discussion of the possibility to reduce the gap value in this way).

\subsection{Large spins}

It is well known that $T_N\sim S^2J$ whereas $D\sim SJ$. Then, the temperature can lie in the range $T_N\gg T\gg D$ for large enough $S$. Besides, quantum corrections to observables decrease upon $S$ increasing and they die out in the limit of $S\to\infty$. In contrast, ratios of temperature corrections to the bare values of observables contain powers of $T/S^2J\sim T/T_N$ so that temperature corrections remain finite in the limit of classical spins (see, e.g., Ref.~\cite{syromyat2d} for detail discussion of this point). As a result, to calculate the gap at $T_N\gg T\gg D$ and $S\gg1$, one can replace ${\cal N}_{\bf q}$ by $T/\epsilon_{\bf q}$ in Eqs.~\eqref{o4} and \eqref{o3} and discard $T$-independent terms. We obtain in this way for the gap instead of Eq.~\eqref{gap}
\begin{equation}
\Delta_\gg^2 = C_\gg S^2\omega_0^2\frac{T}{S^2J},
\label{gap_cl}
\end{equation}
where the constant $C_\gg$ is defined as (cf.\ Eq.~\eqref{c}) 
\begin{equation}
\label{c>>}
C_\gg = \frac{3J^2}{\omega_0^2}\frac1N\sum_\q \frac{(J_\0-J_\q)^2((Q_\q^{xx}-Q_\q^{yy})^2-4(Q_\q^{xy})^2)}{(J_\0^2-J_\q^2)^{2}} \approx 0.018.
\end{equation}
The summation over $q\sim1$ gives the main contribution in Eq.~\eqref{c>>}.

Damping estimation leads to Eqs.~\eqref{g+tn01} and \eqref{g-tn01} for $\sin\theta_\kk\sqrt{\frac{\omega_0}{J}}\gg k\gg \frac{\Delta_{\gg}}{D}$ and $T_N\gg T\gg D$, where now $B=0$ and $A$ is a constant of the order of unity. These expressions give for ratios $\Gamma_{\kk}^\pm/\epsilon_\kk^\pm \sim \omega_0^2T/k^2JD^2$. Then, one obtains using Eq.~\eqref{gap_cl} near the peak position at $k\sim \frac{\Delta_{\gg}}{D}$ and at fixed $\varphi_\kk$ and $\theta_\kk$
\begin{equation}
	\frac{\Gamma_{\kk}^\pm}{\epsilon_\kk^\pm} \sim \rm const.
\end{equation}
Thus, we demonstrate a breakdown of a small fraction of long-wavelength magnons (with $k\sim \frac{\Delta_{\gg}}{D}$ and $\sin2\theta_\kk\sim1$) for $S\gg1$ at small temperature $T\ll T_N$.

\subsection{Competing magnetocrystalline anisotropy}
\label{companis}

We assume now that the gap $\Delta$ in the spectrum \eqref{gap} is decreased by the competing magnetocrystalline anisotropy so that $\Delta\ll \sqrt S\omega_0$. Counterparts of Eqs.~\eqref{g+tn01} and \eqref{g-tn01} which are valid at 
$
\frac{\Delta}{S\sqrt{\omega_0}\sin\theta_\kk}
\gg
k
\gg
\frac{\Delta}{D}
$,
$\sin\theta_\kk \gg \sqrt{\frac{\Delta}{S\omega_0}}$,
and $D \gg T \gg \frac{\Delta^2}{S^2\omega_0k\sin^2\theta_\kk}$ have the form
\begin{eqnarray}
\label{g+tn0comp}
\Gamma_{\kk}^+ &=& 
\frac{\omega_0^2}{kJ}
\left(\frac TD\right)^4
\frac{\pi^3(1+12 c_{xy})^2}{5760} 
\left(
1-
\frac{ f(\varphi_\kk)}{4(1+18 c_{xx}+6 c_{xy})}
\right)
\sin^2 2\theta_\kk
\nonumber\\
&&
\approx
\frac{\omega_0^2}{kJ}
\left(\frac TD\right)^4
6.1\cdot10^{-4}
(1-0.16f(\varphi_\kk))
\sin^2 2\theta_\kk,\\
\label{g-tn0comp}
\Gamma_{\kk}^- &=& 
\Gamma_{\kk}^+
\frac{f(\varphi_\kk)}{4(1+18 c_{xx}+6 c_{xy})-f(\varphi_\kk)}.
\end{eqnarray}
One concludes from these equations that the damping becomes of the order of the real part of the spectrum near the peak position at $k\sim \Delta/D$ if the gap value is decreased so that the inequality $\Delta\alt\omega_0(T/D)^2\ll\omega_0$ holds.

We expect that there is a small chance of success to find a cubic AF in which the magnetocrystalline anisotropy cancels almost completely the dipolar gap. For instance, the anisotropy in the most perfectly isotropic cubic AFs $\rm TlMnF_3$ (Ref.~\cite{TlMn}) and $\rm RbMnF_3$ (Refs.~\cite{rbmnf3,rbmnf3_2,rbmnf3_3}) competes with the dipolar one being of the same order of magnitude. But it turns out to be slightly greater in these substances than the dipolar anisotropy so that the easy directions are space diagonals of the cube. The resultant gaps in these materials turn out to be even slightly greater than the dipolar gap given by Eq.~\eqref{gap}.

However a way was proposed to change gradually values of the anisotropy (and the gap) in $\rm TlMnF_3$ (Ref.~\cite{tlmnf3}) and $\rm RbMnF_3$ (Ref.~\cite{rbmnf3co}) by replacing a tiny amount of $\rm Mn^{2+}$ ions by $\rm Co^{2+}$. As $\rm Mn^{2+}$ ions are in spherically symmetric states with $L=0$ and $S=5/2$ in these compounds, the magnetocrystalline anisotropy is tiny so that the anisotropy field favoring $\langle111\rangle$ directions is equal to several oersteds only. In contrast, $\rm Co^{2+}$ ions have $L\ne0$ and $S=3/2$. As a consequence, the effect of spin-orbit interaction is much more pronounced: the anisotropy field selecting $\langle100\rangle$ direction is four orders of magnitude larger than that of $\rm Mn^{2+}$. Thus, two single-ion anisotropies on $\rm Mn^{2+}$ and $\rm Co^{2+}$ ions compete in mixed compounds ${\rm TlMn}_{1-x}{\rm Co}_x{\rm F}_3$ and ${\rm RbMn}_{1-x}{\rm Co}_x{\rm F}_3$. Due to the great difference between the anisotropies magnitudes on $\rm Mn^{2+}$ and $\rm Co^{2+}$, a very small $x=x_*$ is required to change the easy direction of the whole sample from $\langle111\rangle$ to $\langle100\rangle$: $x_*\approx0.0004$ and $x_*\approx0.00034$ for ${\rm TlMn}_{1-x}{\rm Co}_x{\rm F}_3$ and ${\rm RbMn}_{1-x}{\rm Co}_x{\rm F}_3$, respectively. The gap value is reduced considerably at $x\approx x_*$. 

It has been shown recently \cite{utesov} that states near the magnon band edges can become localized in disordered systems with gapped spectrum. However, according to estimations made in Appendix~\ref{defect}, states with $k\gg x (A'/J)^2\sim10^{-6}$~\AA$^{-1}$, where $A'$ is the value of the magnetocrystalline anisotropy on $\rm Co^{2+}$, remain propagating in materials under consideration. Besides, the magnon damping due to the scattering on impurities is negligible at such $\kk$. On the other hand, Eqs.~\eqref{g+tn0comp} and \eqref{g-tn0comp} predict the magnon breakdown at $k\lesssim 10^{-3}\div10^{-4}$~\AA$^{-1}$ due to the magnon interaction with each other if the gap is reduced considerably. Notice also that momenta of summation $q\lesssim 10^{-6}$~\AA$^{-1}$ are inessential in the calculations leading to Eqs.~\eqref{g+tn0comp} and \eqref{g-tn0comp}. Then, ${\rm TlMn}_{1-x}{\rm Co}_x{\rm F}_3$ and ${\rm RbMn}_{1-x}{\rm Co}_x{\rm F}_3$ can be suitable for testing of our predictions. 

Unfortunately, it would be difficult to carry out corresponding experiments because the characteristic values of momenta of overdamped magnons are quite small being of the order of $10^{-3}\div10^{-4}$~\AA$^{-1}$. However, bearing in mind recent progress in neutron spin-echo technique \cite{bayr,mesot}, we hope that the corresponding measurements will become feasible in the near future. It is also possible that more suitable substances will be found which have larger values of momenta at which the discussed anomalies arise in the damping. 

It should be noted also that our conclusion about suitability of the mixed compounds for the observation of the magnon breakdown is based on estimations made in Appendix~\ref{defect} in the first order in $x$. As soon as defects change the bare spectrum considerably at $x\approx x_*$ and $k\ll1$, these estimations must be used with caution. In particular, one cannot fully exclude the possibility of great spectrum change by terms of higher orders in $x$. It is difficult to analyze the whole series in $x$ but we point out that all the expected contributions are small as compared to those of the first order in $x$ due to the smallness of all kinds of anisotropy in comparison with the exchange constant.

\section{Conclusion}
\label{conc}

To conclude, we discuss magnon damping in Heisenberg AF on a simple cubic lattice with dipolar forces at small temperature $T\ll T_N$. In accordance with previous results, it is demonstrated that dipolar forces split the magnon spectrum into two branches. The classical gapless spectra of long-wavelength magnons into two branches are given by Eq.~\eqref{spectr}. It is found that quantum and thermal fluctuations modify the spectrum considerably near points $\kk=\bf 0$ and $\kk=\kk_0$: the gap $\Delta$ (see Eq.~\eqref{gap}) appears in the spectrum. The gap is accompanied by anisotropic corrections to the ground state energy \eqref{de} which make cube edges easy directions for the staggered magnetization. These effects are of ``order-by-disorder'' nature. The renormalized spectrum of long-wavelength magnons is given by Eq.~\eqref{specren}. 

It is shown that magnons are well defined quasiparticles for all $\kk$ at $S\sim1$ and the ratio $\Gamma_\kk^\pm/\epsilon_\kk^\pm$ has the peak at $k\sim \Delta/D$ which height is proportional to $T/D\ll1$ (see Fig.~\ref{zatuch}). We discuss some possibilities of observing a phenomenon contradicting expectation of the quasiparticle concept: the breakdown of some part of long-wavelength magnons. In particular, it is shown that $\Gamma_\kk^\pm/\epsilon_\kk^\pm\sim{\rm const}$ (at fixed $\varphi_\kk$ and $\theta_\kk$) near the peak position when $S\gg1$ and $T_N/S\ll T\ll T_N$. It is also shown that a single-ion anisotropy which competes with the dipolar one \eqref{de} reduces the gap value enhancing the peak height. The peak height can reach a value of the order of unity for sufficiently small gap that signifies the breakdown of long-wavelength magnons with momenta lying near the peak position. The gap can be decreased and the magnon breakdown can be stimulated also by replacing of a small amount of magnetic atoms by those with single-ion anisotropy competing with the dipolar one. We argue that this effect can be observed in ${\rm TlMn}_{1-x}{\rm Co}_x{\rm F}_3$ and ${\rm RbMn}_{1-x}{\rm Co}_x{\rm F}_3$ at $x\sim0.0004$.

\begin{acknowledgments}

This work is supported by Russian Scientific Fund Grant No.\ 14-22-00281.

\end{acknowledgments}

\appendix

\section{Green's functions and general expression for the spectrum}
\label{ap_spectr}

Solution of Eq.~\eqref{Dyson} has the following form in the spin-wave approximation (i.e., with zero self-energy parts):
\begin{eqnarray}
G(\omega,\kk) &=& 
\frac
{\left(\omega+E_\kk\right) \left(\omega^2+B_{\kk+\kk_0}^2-E_{\kk+\kk_0}^2\right)+2 {\cal B}_\kk {\cal E}_\kk B_{\kk+\kk_0} + {\cal E}_\kk^2\left(\omega-E_{\kk+\kk_0}\right) -{\cal B}_\kk^2 \left(\omega+E_{\kk+\kk_0}\right)}
{\left(\omega^2-\left(\epsilon_{0\kk}^+\right)^2\right)\left(\omega^2-\left(\epsilon_{0\kk}^-\right)^2\right)},\nonumber\\
F^\dagger(\omega,\kk) &=& 
\frac
{-B_\kk \left(\omega^2 + B_{\kk+\kk_0}^2 - E_{\kk+\kk_0}^2\right)-2 {\cal B}_\kk {\cal E}_\kk E_{\kk+\kk_0}+B_{\kk+\kk_0} \left({\cal B}_\kk^2+{\cal E}_\kk^2\right)}
{\left(\omega^2-\left(\epsilon_{0\kk}^+\right)^2\right)\left(\omega^2-\left(\epsilon_{0\kk}^-\right)^2\right)},
\label{gfsw}\\
{\cal F}^\dagger(\omega,\kk) &=& 
\frac
{- {\cal E}_\kk\left(B_{\kk+\kk_0} \left(\omega+E_\kk\right)+B_\kk \left(\omega-E_{\kk+\kk_0}\right)\right)-{\cal B}_\kk \left(B_\kk B_{\kk+\kk_0}+\left(\omega+E_\kk\right) \left(\omega-E_{\kk+\kk_0}\right)+{\cal E}_\kk^2-{\cal B}_\kk^2\right)}
{\left(\omega^2-\left(\epsilon_{0\kk}^+\right)^2\right)\left(\omega^2-\left(\epsilon_{0\kk}^-\right)^2\right)},\nonumber\\
{\cal G}(\omega,\kk) &=& 
\frac
{{\cal B}_\kk \left(B_\kk \left(\omega+E_{\kk+\kk_0}\right)-B_{\kk+\kk_0} \left(\omega+E_\kk\right)\right)+ {\cal E}_\kk\left(-B_\kk B_{\kk+\kk_0}+\left(\omega+E_\kk\right) \left(\omega+E_{\kk+\kk_0}\right)+{\cal E}_\kk^2-{\cal B}_\kk^2\right)}
{\left(\omega^2-\left(\epsilon_{0\kk}^+\right)^2\right)\left(\omega^2-\left(\epsilon_{0\kk}^-\right)^2\right)},\nonumber
\end{eqnarray}
where energies $\epsilon_{0\kk}^\pm$ of the two magnon branches have the form \eqref{specsw}. By setting $\omega_0=0$, one leads from Eqs.~\eqref{gfsw} to the Green's functions of Heisenberg AF (see, e.g., Ref.~\cite{af2d}): ${\cal F}^\dagger(\omega,\kk)={\cal G}(\omega,\kk)=0$, $G(\omega,\kk)=G_c(\omega,\kk)$, and $F^\dagger(\omega,\kk)=F_c^\dagger(\omega,\kk)$, where
\begin{eqnarray}
\label{gc}
G_c(\omega,\kk) &=& \frac{SJ_{\bf 0}+\omega}{\omega^2-S^2(J_{\bf 0}^2-J_{\bf k}^2)},\\
\label{fc}
F_c^\dagger(\omega,\kk) &=& -\frac{SJ_{\bf k}}{\omega^2-S^2(J_{\bf 0}^2-J_{\bf k}^2)}.
\end{eqnarray}

The explicit expression for $\Omega(\omega,\kk)$ introduced in Eq.~\eqref{d1} has the form
\begin{equation}
\label{omega}
\Omega(\omega,\kk) = 
\delta\Omega(\omega,\kk) + \delta\Omega(\omega,\kko),
\end{equation}
where
\begin{eqnarray}
\label{domega}
\delta\Omega(\omega,\kk) &=&
\left(B_\kk \left(\omega^2+B_\kko^2-E_\kko^2\right)+2 {\cal B}_\kk {\cal E}_\kk E_\kko-B_\kko \left({\cal B}_\kk^2+{\cal E}_\kk^2\right)\right) 
\left(\Pi_\kk+\Pi^\dagger_\kk\right)\nonumber\\
&-&
\left(E_\kk \left(\omega^2+B_\kko^2-E_\kko^2\right)+2{\cal B}_\kk {\cal E}_\kk B_\kko -  E_\kko\left({\cal B}_\kk^2 + {\cal E}_\kk^2\right)\right) 
\left(\Sigma_\kk+\overline\Sigma_\kk\right)\nonumber\\
&+&
\omega \left(\omega^2+B_\kko^2-E_\kko^2+{\cal E}_\kk^2-{\cal B}_\kk^2\right) 
\left(\Sigma_\kk-\overline\Sigma_\kk\right)\nonumber\\
&+&
\omega \left({\cal B}_\kk \left(E_\kk-E_\kko\right) + {\cal E}_\kk\left(B_\kk+B_\kko\right)\right) 
\left({\cal P}_\kk+{\cal P}^\dagger_\kk\right)\nonumber\\
&-&
\omega \left({\cal B}_\kk \left(B_\kk-B_\kko\right) + {\cal E}_\kk\left(E_\kk+E_\kko\right)\right) 
\left({\cal S}_\kk+\overline{\cal S}_\kk\right)\nonumber\\
&+&
\left( 
{\cal E}_\kk \left(B_\kko E_\kk-B_\kk E_\kko\right) 
+{\cal B}_\kk \left(\omega^2+B_\kk B_\kko-E_\kk E_\kko + {\cal E}_\kk^2-{\cal B}_\kk^2 \right)
\right) 
\left({\cal P}_\kk-{\cal P}^\dagger_\kk\right)\nonumber\\
&+&
\left(
{\cal B}_\kk \left(B_\kk E_\kko-B_\kko E_\kk\right) 
+{\cal E}_\kk \left(\omega^2-B_\kk B_\kko +E_\kk E_\kko + {\cal E}_\kk^2 - {\cal B}_\kk^2\right)
\right) 
\left({\cal S}_\kk-\overline{\cal S}_\kk\right).
\end{eqnarray}

One obtains for the first $1/S$ corrections to $\Omega\left(\omega=\epsilon_{0\kk}^\pm+i\delta,\kk\right)$ at $k\ll1$ in the leading order in $k$ and $\omega_0/J$
\begin{subequations}
\label{omega1}
\begin{eqnarray}
\mp\frac{\Omega\left(\omega=\epsilon_{0\kk}^\pm+i\delta,\kk\right)}{2\sqrt{d_\kk}}
&=&
\frac{SJ_{\bf 0}}{2} \left( -1 \mp (1+36c_{xx})\frac{\cos(2\varphi_\kk)}{L_1(\varphi_\kk)}\right)
\left(\Pi_\kk+\Pi^\dagger_\kk-\Sigma_\kk-\overline{\Sigma}_\kk\right)\\
&&{}+
\frac{SJ_{\bf 0}}{2} \left( 1 \mp (1+36c_{xx})\frac{\cos(2\varphi_\kk)}{L_1(\varphi_\kk)}\right)
\left(\Pi_\kko+\Pi^\dagger_\kko+\Sigma_\kko+\overline{\Sigma}_\kko\right)\\
&&{}\pm
i SJ_{\bf 0}\frac{1+12c_{xy}}{2} \frac{\sin(2\varphi_\kk)}{L_1(\varphi_\kk)} 
\left({\cal P}_\kk-{\cal P}_\kk^\dagger+\overline{\cal S}_\kk-{\cal S}_\kk 
+ {\cal P}_\kko-{\cal P}_\kko^\dagger+{\cal S}_\kko-\overline{\cal S}_\kko\right)\\
&&{}
- k \frac{SJ_{\bf 0}}{2\sqrt{3}} 
\left( \Sigma_\kk - \overline{\Sigma}_\kk + \Sigma_\kko - \overline{\Sigma}_\kko \right)\\
&&{} 
\mp
k SJ_{\bf 0}\frac{1+36c_{xx}}{2\sqrt{3}} 
\frac{\cos(2\varphi_\kk)}{L_1(\varphi_\kk)}
\left( \Sigma_\kk - \overline{\Sigma}_\kk - \Sigma_\kko + \overline{\Sigma}_\kko \right)\\
&&{}\pm
i k SJ_{\bf 0}
\frac{1+12c_{xy}}{2\sqrt{3}} \frac{\sin(2\varphi_\kk)}{L_1(\varphi_\kk)} 
\left( {\cal S}_\kk + \overline{\cal S}_\kk - {\cal S}_\kko - \overline{\cal S}_\kko \right)\\
&&{}+
k^2\frac{SJ_{\bf 0}}{24} 
\left( 1 \pm (1+36c_{xx})\frac{\cos(2\varphi_\kk)}{L_1(\varphi_\kk)}\right)
\left(\Pi_\kk+\Pi^\dagger_\kk+\Sigma_\kk+\overline{\Sigma}_\kk\right)\\
&&{}+
k^2\frac{SJ_{\bf 0}}{24} \left( -1 \pm (1+36c_{xx})\frac{\cos(2\varphi_\kk)}{L_1(\varphi_\kk)}\right)
\left(\Pi_\kko+\Pi^\dagger_\kko-\Sigma_\kko-\overline{\Sigma}_\kko\right)\\
&&{}\mp
i k^2SJ_{\bf 0}\frac{1+12c_{xy}}{24} \frac{\sin(2\varphi_\kk)}{L_1(\varphi_\kk)} 
\left({\cal P}_\kk-{\cal P}_\kk^\dagger-\overline{\cal S}_\kk+{\cal S}_\kk 
+ {\cal P}_\kko-{\cal P}_\kko^\dagger-{\cal S}_\kko+\overline{\cal S}_\kko\right),
\end{eqnarray}
\end{subequations}
where self-energy parts are taken at $\omega=\epsilon_{1\kk}^\pm+i\delta$. All terms in Eq.~\eqref{omega1} contribute to results presented in the main text for the damping due to decay processes, while only terms \eqref{omega1}(a)--(c) give the leading contributions to the damping due to confluence processes. Expressions for combinations of self-energy parts which arise in Eq.~\eqref{omega1} are presented in Appendix~\ref{expsep}.

\section{Expressions for self-energy parts}
\label{expsep}

In this appendix, we present expressions for some combinations of self-energy parts which arise in Eq.~\eqref{omega1}. Only contributions are shown below which are of the first order in $1/S$ and which originate from the loop diagram depicted in Fig.~\ref{dia}(b). To make all expressions more compact, we move arguments of Green's functions to subscripts and introduce the following notation: $k=(\omega,\kk)$, $q=(\omega_q,{\bf q})$, and $k_0=(0,\ko)$.
\begin{eqnarray}
&&\Pi_\kk+\Pi^\dagger_\kk-\Sigma_\kk-\overline{\Sigma}_\kk = \\
&&-\frac{S}{2N} T\sum_{\omega_q,\bf q}
\left(
({\cal F}_{k-q}-{\cal F}^\dagger_{k-q}+{\cal G}_{k-q}-\overline{\cal G}_{k-q}) ({\cal F}_{k_0+q}-{\cal F}^\dagger_{k_0+q}-{\cal G}_{k_0+q}+\overline{\cal G}_{k_0+q}) Q^{xz}_{\kk-\q} Q^{xz}_\q
\right.\nonumber\\
&&+(F_{k+k_0-q}-F^\dagger_{k+k_0-q}+G_{k+k_0-q}-\overline{G}_{k+k_0-q}) (F_q-F^\dagger_q-G_q+\overline{G}_q) Q^{xz}_{\kk+\ko-\q} Q^{xz}_\q\nonumber\\
&&+(F_{k+k_0-q}+F^\dagger_{k+k_0-q}-G_{k+k_0-q}-\overline{G}_{k+k_0-q}) (F_q+F^\dagger_q+G_q+\overline{G}_q) Q^{xz2}_\q\nonumber\\
&&+({\cal F}_{k-q}+{\cal F}^\dagger_{k-q}-{\cal G}_{k-q}-\overline{\cal G}_{k-q}) ({\cal F}_{k_0+q}+{\cal F}^\dagger_{k_0+q}+{\cal G}_{k_0+q}+\overline{\cal G}_{k_0+q}) Q^{xz}_\q Q^{xz}_{\ko+\q}\nonumber\\
&&-2 i (F_{k-q}+G_{k-q}) ({\cal F}^\dagger_{k_0+q}-\overline{\cal G}_{k_0+q}) Q^{xz}_{\kk-\q} Q^{yz}_{\kk+\ko}-2 i ({\cal F}_{k+k_0-q}+{\cal G}_{k+k_0-q}) (F^\dagger_q-\overline{G}_q) Q^{xz}_{\kk+\ko-\q} Q^{yz}_{\kk+\ko}\nonumber\\
&&+2 i (F^\dagger_q+G_q) ({\cal F}_{k-q}-\overline{\cal G}_{k-q}) Q^{xz}_\q Q^{yz}_{\kk+\ko}-2 i (F^\dagger_{k+k_0-q}-G_{k+k_0-q}) ({\cal F}_{k_0+q}+\overline{\cal G}_{k_0+q}) Q^{xz}_\q Q^{yz}_{\kk+\ko}\nonumber\\
&&+2 i ({\cal F}^\dagger_{k_0+q}+{\cal G}_{k_0+q}) (F_{k+k_0-q}-\overline{G}_{k+k_0-q}) Q^{xz}_\q Q^{yz}_{\kk+\ko}-2 i ({\cal F}^\dagger_{k-q}-{\cal G}_{k-q}) (F_q+\overline{G}_q) Q^{xz}_\q Q^{yz}_{\kk+\ko}\nonumber\\
&&+2 i ({\cal F}_{k+k_0-q}-{\cal G}_{k+k_0-q}) (F^\dagger_q+\overline{G}_q) Q^{xz}_\q Q^{yz}_{\kk+\ko}+2 i (F_{k-q}-G_{k-q}) ({\cal F}^\dagger_{k_0+q}+\overline{\cal G}_{k_0+q}) Q^{xz}_{\ko+\q} Q^{yz}_{\kk+\ko}\nonumber\\
&&
-4 ({\cal F}_{k+k_0-q} {\cal F}^\dagger_{k_0+q} 
+ F_{k-q} F^\dagger_q 
+ {\cal G}_{k+k_0-q} \overline{\cal G}_{k_0+q} 
+ G_{k-q} \overline{G}_q) Q^{yz2}_{\kk+\ko}\nonumber\\
&&+i (-{\cal F}_{k_0+q}+{\cal F}^\dagger_{k_0+q}+{\cal G}_{k_0+q}-\overline{\cal G}_{k_0+q}) (F_{k+k_0-q}+F^\dagger_{k+k_0-q}-G_{k+k_0-q}-\overline{G}_{k+k_0-q}) Q^{xz}_\q Q^{yz}_{\kk-\q}\nonumber\\
&&+2 (-{\cal F}_{k+k_0-q}+{\cal G}_{k+k_0-q}) ({\cal F}^\dagger_{k_0+q}-\overline{\cal G}_{k_0+q}) Q^{yz}_{\kk+\ko} Q^{yz}_{\kk-\q}\nonumber\\
&&+i ({\cal F}_{k-q}+{\cal F}^\dagger_{k-q}-{\cal G}_{k-q}-\overline{\cal G}_{k-q}) (-F_q+F^\dagger_q+G_q-\overline{G}_q) Q^{xz}_\q Q^{yz}_{\kk+\ko-\q}\nonumber\\
&&+2 (-F_{k-q}+G_{k-q}) (F^\dagger_q-\overline{G}_q) Q^{yz}_{\kk+\ko} Q^{yz}_{\kk+\ko-\q}+2 (-{\cal F}_{k+k_0-q}+{\cal G}_{k+k_0-q}) ({\cal F}^\dagger_{k_0+q}-\overline{\cal G}_{k_0+q}) Q^{yz}_{\kk+\ko} Q^{yz}_\q\nonumber\\
&&+i (-{\cal F}_{k_0+q}+{\cal F}^\dagger_{k_0+q}+{\cal G}_{k_0+q}-\overline{\cal G}_{k_0+q}) (F_{k+k_0-q}+F^\dagger_{k+k_0-q}-G_{k+k_0-q}-\overline{G}_{k+k_0-q}) Q^{xz}_\q Q^{yz}_\q\nonumber\\
&&+i ({\cal F}_{k_0+q}+{\cal F}^\dagger_{k_0+q}-{\cal G}_{k_0+q}-\overline{\cal G}_{k_0+q}) (-F_{k-q}+F^\dagger_{k-q}-G_{k-q}+\overline{G}_{k-q}) Q^{xz}_{\kk-\q} Q^{yz}_{\ko+\q}\nonumber\\
&&+i (-{\cal F}_{k+k_0-q}+{\cal F}^\dagger_{k+k_0-q}-{\cal G}_{k+k_0-q}+\overline{\cal G}_{k+k_0-q}) (F_q+F^\dagger_q-G_q-\overline{G}_q) Q^{xz}_{\kk+\ko-\q} Q^{yz}_{\ko+\q}\nonumber\\
&&+i ({\cal F}_{k-q}+{\cal F}^\dagger_{k-q}-{\cal G}_{k-q}-\overline{\cal G}_{k-q}) (-F_q+F^\dagger_q+G_q-\overline{G}_q) Q^{xz}_\q Q^{yz}_{\ko+\q}\nonumber\\
&&+i ({\cal F}_{k+k_0-q}+{\cal F}^\dagger_{k+k_0-q}-{\cal G}_{k+k_0-q}-\overline{\cal G}_{k+k_0-q}) (-F_q+F^\dagger_q-G_q+\overline{G}_q) Q^{xz}_\q Q^{yz}_{\ko+\q}\nonumber\\
&&+i (-{\cal F}_{k_0+q}+{\cal F}^\dagger_{k_0+q}-{\cal G}_{k_0+q}+\overline{\cal G}_{k_0+q}) (F_{k-q}+F^\dagger_{k-q}-G_{k-q}-\overline{G}_{k-q}) Q^{xz}_{\ko+\q} Q^{yz}_{\ko+\q}\nonumber\\
&&+2 ({\cal F}^\dagger_{k_0+q}-{\cal G}_{k_0+q}) (-{\cal F}_{k+k_0-q}+\overline{\cal G}_{k+k_0-q}) Q^{yz}_{\kk+\ko} Q^{yz}_{\ko+\q}\nonumber\\
&&+2 ({\cal F}^\dagger_{k+k_0-q}-{\cal G}_{k+k_0-q}) (-{\cal F}_{k_0+q}+\overline{\cal G}_{k_0+q}) Q^{yz}_{\kk+\ko} Q^{yz}_{\ko+\q}
+2 (F^\dagger_{k-q}-G_{k-q}) (-F_q+\overline{G}_q) Q^{yz}_{\kk+\ko} Q^{yz}_{\ko+\q}\nonumber\\
&&+2 (F^\dagger_q-G_q) (-F_{k-q}+\overline{G}_{k-q}) Q^{yz}_{\kk+\ko} Q^{yz}_{\ko+\q}+2 (-F_{k-q}+G_{k-q}) (F^\dagger_q-\overline{G}_q) Q^{yz}_{\kk+\ko} Q^{yz}_{\ko+\q}\nonumber\\
&&-({\cal F}_{k+k_0-q}+{\cal F}^\dagger_{k+k_0-q}-{\cal G}_{k+k_0-q}-\overline{\cal G}_{k+k_0-q}) ({\cal F}_{k_0+q}+{\cal F}^\dagger_{k_0+q}-{\cal G}_{k_0+q}-\overline{\cal G}_{k_0+q}) Q^{yz}_{\kk-\q} Q^{yz}_{\ko+\q}\nonumber\\
&&-(F_{k-q}+F^\dagger_{k-q}-G_{k-q}-\overline{G}_{k-q}) (F_q+F^\dagger_q-G_q-\overline{G}_q) Q^{yz}_{\kk+\ko-\q} Q^{yz}_{\ko+\q}\nonumber\\
&&-({\cal F}_{k+k_0-q}+{\cal F}^\dagger_{k+k_0-q}-{\cal G}_{k+k_0-q}-\overline{\cal G}_{k+k_0-q}) ({\cal F}_{k_0+q}+{\cal F}^\dagger_{k_0+q}-{\cal G}_{k_0+q}-\overline{\cal G}_{k_0+q}) Q^{yz}_\q Q^{yz}_{\ko+\q}\nonumber\\
&&\left.-(F_{k-q}+F^\dagger_{k-q}-G_{k-q}-\overline{G}_{k-q}) (F_q+F^\dagger_q-G_q-\overline{G}_q) Q^{yz2}_{\ko+\q}
\right).\nonumber
\end{eqnarray} 

\begin{eqnarray}
&&\Pi_\kk+\Pi^\dagger_\kk+\Sigma_\kk+\overline{\Sigma}_\kk = \\
&&-\frac{S}{2N} T\sum_{\omega_q,\bf q}
\left(
4 ({\cal F}_{k-q} {\cal F}^\dagger_{k_0+q}
+F_{k+k_0-q} F^\dagger_q 
+{\cal G}_{k-q} \overline{\cal G}_{k_0+q}
+G_{k+k_0-q} \overline{G}_q) Q^{xz2}_\kk
\right.\nonumber\\
&&+2 ({\cal F}_{k-q}+{\cal G}_{k-q}) ({\cal F}^\dagger_{k_0+q}+\overline{\cal G}_{k_0+q}) Q^{xz}_\kk Q^{xz}_{\kk-\q}+2 (F_{k+k_0-q}+G_{k+k_0-q}) (F^\dagger_q+\overline{G}_q) Q^{xz}_\kk Q^{xz}_{\kk+\ko-\q}\nonumber\\
&&+2 ({\cal F}^\dagger_{k_0+q}+{\cal G}_{k_0+q}) ({\cal F}_{k-q}+\overline{\cal G}_{k-q}) Q^{xz}_\kk Q^{xz}_\q+2 ({\cal F}^\dagger_{k-q}+{\cal G}_{k-q}) ({\cal F}_{k_0+q}+\overline{\cal G}_{k_0+q}) Q^{xz}_\kk Q^{xz}_\q\nonumber\\
&&+2 (F^\dagger_q+G_q) (F_{k+k_0-q}+\overline{G}_{k+k_0-q}) Q^{xz}_\kk Q^{xz}_\q+2 (F^\dagger_{k+k_0-q}+G_{k+k_0-q}) (F_q+\overline{G}_q) Q^{xz}_\kk Q^{xz}_\q
\nonumber\\
&&+2 (F_{k+k_0-q}+G_{k+k_0-q}) (F^\dagger_q+\overline{G}_q) Q^{xz}_\kk Q^{xz}_\q
+2 ({\cal F}_{k-q}+{\cal G}_{k-q}) ({\cal F}^\dagger_{k_0+q}+\overline{\cal G}_{k_0+q}) Q^{xz}_\kk Q^{xz}_{\ko+\q}\nonumber\\
&&+({\cal F}_{k-q}+{\cal F}^\dagger_{k-q}+{\cal G}_{k-q}+\overline{\cal G}_{k-q}) ({\cal F}_{k_0+q}+{\cal F}^\dagger_{k_0+q}+{\cal G}_{k_0+q}+\overline{\cal G}_{k_0+q}) Q^{xz}_{\kk-\q} Q^{xz}_\q\nonumber\\
&&+(F_{k+k_0-q}+F^\dagger_{k+k_0-q}+G_{k+k_0-q}+\overline{G}_{k+k_0-q}) (F_q+F^\dagger_q+G_q+\overline{G}_q) Q^{xz}_{\kk+\ko-\q} Q^{xz}_\q\nonumber\\
&&+(F_{k+k_0-q}+F^\dagger_{k+k_0-q}+G_{k+k_0-q}+\overline{G}_{k+k_0-q}) (F_q+F^\dagger_q+G_q+\overline{G}_q) Q^{xz2}_\q\nonumber\\
&&+({\cal F}_{k-q}+{\cal F}^\dagger_{k-q}+{\cal G}_{k-q}+\overline{\cal G}_{k-q}) ({\cal F}_{k_0+q}+{\cal F}^\dagger_{k_0+q}+{\cal G}_{k_0+q}+\overline{\cal G}_{k_0+q}) Q^{xz}_\q Q^{xz}_{\ko+\q}\nonumber\\
&&+2 i (-F_{k+k_0-q}+G_{k+k_0-q}) ({\cal F}^\dagger_{k_0+q}+\overline{\cal G}_{k_0+q}) Q^{xz}_\kk Q^{yz}_{\kk-\q}+2 i (-{\cal F}_{k-q}+{\cal G}_{k-q}) (F^\dagger_q+\overline{G}_q) Q^{xz}_\kk Q^{yz}_{\kk+\ko-\q}\nonumber\\
&&+i ({\cal F}_{k_0+q}+{\cal F}^\dagger_{k_0+q}+{\cal G}_{k_0+q}+\overline{\cal G}_{k_0+q}) (-F_{k+k_0-q}+F^\dagger_{k+k_0-q}+G_{k+k_0-q}-\overline{G}_{k+k_0-q}) Q^{xz}_\q Q^{yz}_{\kk-\q}\nonumber\\
&&+i (-{\cal F}_{k-q}+{\cal F}^\dagger_{k-q}+{\cal G}_{k-q}-\overline{\cal G}_{k-q}) (F_q+F^\dagger_q+G_q+\overline{G}_q) Q^{xz}_\q Q^{yz}_{\kk+\ko-\q}\nonumber\\
&&+i (-{\cal F}_{k_0+q}+{\cal F}^\dagger_{k_0+q}+{\cal G}_{k_0+q}-\overline{\cal G}_{k_0+q}) (F_{k+k_0-q}+F^\dagger_{k+k_0-q}+G_{k+k_0-q}+\overline{G}_{k+k_0-q}) Q^{xz}_\q Q^{yz}_\q\nonumber\\
&&+2 i (F^\dagger_q-G_q) ({\cal F}_{k+k_0-q}+\overline{\cal G}_{k+k_0-q}) Q^{xz}_\kk Q^{yz}_{\ko+\q}+2 i (F^\dagger_{k-q}+G_{k-q}) (-{\cal F}_{k_0+q}+\overline{\cal G}_{k_0+q}) Q^{xz}_\kk Q^{yz}_{\ko+\q}\nonumber\\
&&+2 i ({\cal F}^\dagger_{k_0+q}-{\cal G}_{k_0+q}) (F_{k-q}+\overline{G}_{k-q}) Q^{xz}_\kk Q^{yz}_{\ko+\q}+2 i ({\cal F}_{k-q}+{\cal G}_{k-q}) (F^\dagger_q-\overline{G}_q) Q^{xz}_\kk Q^{yz}_{\ko+\q}\nonumber\\
&&+2 i ({\cal F}^\dagger_{k+k_0-q}+{\cal G}_{k+k_0-q}) (-F_q+\overline{G}_q) Q^{xz}_\kk Q^{yz}_{\ko+\q}+2 i (F_{k+k_0-q}+G_{k+k_0-q}) ({\cal F}^\dagger_{k_0+q}-\overline{\cal G}_{k_0+q}) Q^{xz}_\kk Q^{yz}_\q\nonumber\\
&&+i (-{\cal F}_{k_0+q}+{\cal F}^\dagger_{k_0+q}-{\cal G}_{k_0+q}+\overline{\cal G}_{k_0+q}) (F_{k-q}+F^\dagger_{k-q}+G_{k-q}+\overline{G}_{k-q}) Q^{xz}_{\kk-\q} Q^{yz}_{\ko+\q}\nonumber\\
&&+i ({\cal F}_{k+k_0-q}+{\cal F}^\dagger_{k+k_0-q}+{\cal G}_{k+k_0-q}+\overline{\cal G}_{k+k_0-q}) (-F_q+F^\dagger_q-G_q+\overline{G}_q) Q^{xz}_{\kk+\ko-\q} Q^{yz}_{\ko+\q}\nonumber\\
&&+i ({\cal F}_{k-q}+{\cal F}^\dagger_{k-q}+{\cal G}_{k-q}+\overline{\cal G}_{k-q}) (-F_q+F^\dagger_q+G_q-\overline{G}_q) Q^{xz}_\q Q^{yz}_{\ko+\q}\nonumber\\
&&+i ({\cal F}_{k+k_0-q}+{\cal F}^\dagger_{k+k_0-q}+{\cal G}_{k+k_0-q}+\overline{\cal G}_{k+k_0-q}) (-F_q+F^\dagger_q-G_q+\overline{G}_q) Q^{xz}_\q Q^{yz}_{\ko+\q}\nonumber\\
&&+i (-{\cal F}_{k_0+q}+{\cal F}^\dagger_{k_0+q}-{\cal G}_{k_0+q}+\overline{\cal G}_{k_0+q}) (F_{k-q}+F^\dagger_{k-q}+G_{k-q}+\overline{G}_{k-q}) Q^{xz}_{\ko+\q} Q^{yz}_{\ko+\q}\nonumber\\
&&-({\cal F}_{k+k_0-q}-{\cal F}^\dagger_{k+k_0-q}-{\cal G}_{k+k_0-q}+\overline{\cal G}_{k+k_0-q}) ({\cal F}_{k_0+q}-{\cal F}^\dagger_{k_0+q}+{\cal G}_{k_0+q}-\overline{\cal G}_{k_0+q}) Q^{yz}_{\kk-\q} Q^{yz}_{\ko+\q}\nonumber\\
&&-(F_{k-q}-F^\dagger_{k-q}-G_{k-q}+\overline{G}_{k-q}) (F_q-F^\dagger_q+G_q-\overline{G}_q) Q^{yz}_{\kk+\ko-\q} Q^{yz}_{\ko+\q}\nonumber\\
&&-({\cal F}_{k+k_0-q}+{\cal F}^\dagger_{k+k_0-q}+{\cal G}_{k+k_0-q}+\overline{\cal G}_{k+k_0-q}) ({\cal F}_{k_0+q}+{\cal F}^\dagger_{k_0+q}-{\cal G}_{k_0+q}-\overline{\cal G}_{k_0+q}) Q^{yz}_\q Q^{yz}_{\ko+\q}\nonumber\\
&&\left.-(F_{k-q}+F^\dagger_{k-q}+G_{k-q}+\overline{G}_{k-q}) (F_q+F^\dagger_q-G_q-\overline{G}_q) Q^{yz2}_{\ko+\q}
\right).\nonumber
\end{eqnarray} 

\begin{eqnarray}
&&{\cal P}_\kk-{\cal P}^\dagger_\kk-{\cal S}_\kk+\overline{\cal S}_\kk = \\
&&-\frac{S}{2N} T\sum_{\omega_q,\bf q}
\left(
2 (F^\dagger_q+G_q) ({\cal F}_{k-q}-\overline{\cal G}_{k-q}) Q^{xz}_{\kk+\ko} Q^{xz}_\q-2 (F^\dagger_{k+k_0-q}-G_{k+k_0-q}) ({\cal F}_{k_0+q}+\overline{\cal G}_{k_0+q}) Q^{xz}_{\kk+\ko} Q^{xz}_\q
\right.\nonumber\\
&&+2 ({\cal F}^\dagger_{k_0+q}+{\cal G}_{k_0+q}) (F_{k+k_0-q}-\overline{G}_{k+k_0-q}) Q^{xz}_{\kk+\ko} Q^{xz}_\q-2 ({\cal F}^\dagger_{k-q}-{\cal G}_{k-q}) (F_q+\overline{G}_q) Q^{xz}_{\kk+\ko} Q^{xz}_\q\nonumber\\
&&+({\cal F}_{k-q}-{\cal F}^\dagger_{k-q}+{\cal G}_{k-q}-\overline{\cal G}_{k-q}) (F_q+F^\dagger_q+G_q+\overline{G}_q) Q^{xz}_{\kk-\q} Q^{xz}_\q\nonumber\\
&&+({\cal F}_{k_0+q}+{\cal F}^\dagger_{k_0+q}+{\cal G}_{k_0+q}+\overline{\cal G}_{k_0+q}) (F_{k+k_0-q}-F^\dagger_{k+k_0-q}+G_{k+k_0-q}-\overline{G}_{k+k_0-q}) Q^{xz}_{\kk+\ko-\q} Q^{xz}_\q\nonumber\\
&&+({\cal F}_{k-q}-{\cal F}^\dagger_{k-q}+{\cal G}_{k-q}-\overline{\cal G}_{k-q}) (F_q+F^\dagger_q+G_q+\overline{G}_q) Q^{xz2}_\q\nonumber\\
&&+({\cal F}_{k_0+q}+{\cal F}^\dagger_{k_0+q}+{\cal G}_{k_0+q}+\overline{\cal G}_{k_0+q}) (F_{k+k_0-q}-F^\dagger_{k+k_0-q}+G_{k+k_0-q}-\overline{G}_{k+k_0-q}) Q^{xz}_\q Q^{xz}_{\ko+\q}\nonumber\\
&&
+4 i ({\cal F}_{k+k_0-q} {\cal F}^\dagger_{k_0+q} 
+F_{k-q} F^\dagger_q 
+{\cal G}_{k+k_0-q} \overline{\cal G}_{k_0+q} 
+G_{k-q} \overline{G}_q) Q^{xz}_{\kk+\ko} Q^{yz}_{\kk+\ko}
\nonumber\\
&&+2 i (F_{k-q}+G_{k-q}) (F^\dagger_q+\overline{G}_q) Q^{xz}_{\kk-\q} Q^{yz}_{\kk+\ko}+2 i ({\cal F}_{k+k_0-q}+{\cal G}_{k+k_0-q}) ({\cal F}^\dagger_{k_0+q}+\overline{\cal G}_{k_0+q}) Q^{xz}_{\kk+\ko-\q} Q^{yz}_{\kk+\ko}\nonumber\\
&&+2 i (F_{k-q}+G_{k-q}) (F^\dagger_q+\overline{G}_q) Q^{xz}_\q Q^{yz}_{\kk+\ko}+2 i ({\cal F}_{k+k_0-q}+{\cal G}_{k+k_0-q}) ({\cal F}^\dagger_{k_0+q}+\overline{\cal G}_{k_0+q}) Q^{xz}_{\ko+\q} Q^{yz}_{\kk+\ko}\nonumber\\
&&-i (F_{k+k_0-q}+F^\dagger_{k+k_0-q}-G_{k+k_0-q}-\overline{G}_{k+k_0-q}) (F_q+F^\dagger_q+G_q+\overline{G}_q) Q^{xz}_\q Q^{yz}_{\kk-\q}\nonumber\\
&&+2 ({\cal F}_{k+k_0-q}-{\cal G}_{k+k_0-q}) (F^\dagger_q+\overline{G}_q) Q^{yz}_{\kk+\ko} Q^{yz}_{\kk-\q}
+2 (F_{k-q}-G_{k-q}) ({\cal F}^\dagger_{k_0+q}+\overline{\cal G}_{k_0+q}) Q^{yz}_{\kk+\ko} Q^{yz}_{\kk+\ko-\q}\nonumber\\
&&-i ({\cal F}_{k-q}+{\cal F}^\dagger_{k-q}-{\cal G}_{k-q}-\overline{\cal G}_{k-q}) ({\cal F}_{k_0+q}+{\cal F}^\dagger_{k_0+q}+{\cal G}_{k_0+q}+\overline{\cal G}_{k_0+q}) Q^{xz}_\q Q^{yz}_{\kk+\ko-\q}\nonumber\\
&&-i ({\cal F}_{k-q}-{\cal F}^\dagger_{k-q}+{\cal G}_{k-q}-\overline{\cal G}_{k-q}) ({\cal F}_{k_0+q}-{\cal F}^\dagger_{k_0+q}-{\cal G}_{k_0+q}+\overline{\cal G}_{k_0+q}) Q^{xz}_\q Q^{yz}_\q\nonumber\\
&&-2 (F_{k-q}+G_{k-q}) ({\cal F}^\dagger_{k_0+q}-\overline{\cal G}_{k_0+q}) Q^{yz}_{\kk+\ko} Q^{yz}_\q
-2 ({\cal F}_{k+k_0-q}+{\cal G}_{k+k_0-q}) (F^\dagger_q-\overline{G}_q) Q^{yz}_{\kk+\ko} Q^{yz}_{\ko+\q}\nonumber\\
&&+2 i ({\cal F}^\dagger_{k_0+q}-{\cal G}_{k_0+q}) ({\cal F}_{k+k_0-q}-\overline{\cal G}_{k+k_0-q}) Q^{xz}_{\kk+\ko} Q^{yz}_{\ko+\q}+2 i ({\cal F}^\dagger_{k+k_0-q}-{\cal G}_{k+k_0-q}) ({\cal F}_{k_0+q}-\overline{\cal G}_{k_0+q}) Q^{xz}_{\kk+\ko} Q^{yz}_{\ko+\q}\nonumber\\
&&+2 i (F^\dagger_q-G_q) (F_{k-q}-\overline{G}_{k-q}) Q^{xz}_{\kk+\ko} Q^{yz}_{\ko+\q}+2 i (F^\dagger_{k-q}-G_{k-q}) (F_q-\overline{G}_q) Q^{xz}_{\kk+\ko} Q^{yz}_{\ko+\q}\nonumber\\
&&-i (F_{k-q}-F^\dagger_{k-q}+G_{k-q}-\overline{G}_{k-q}) (F_q-F^\dagger_q+G_q-\overline{G}_q) Q^{xz}_{\kk-\q} Q^{yz}_{\ko+\q}\nonumber\\
&&-i ({\cal F}_{k+k_0-q}-{\cal F}^\dagger_{k+k_0-q}+{\cal G}_{k+k_0-q}-\overline{\cal G}_{k+k_0-q}) ({\cal F}_{k_0+q}-{\cal F}^\dagger_{k_0+q}+{\cal G}_{k_0+q}-\overline{\cal G}_{k_0+q}) Q^{xz}_{\kk+\ko-\q} Q^{yz}_{\ko+\q}\nonumber\\
&&-i (F_{k-q}-F^\dagger_{k-q}+G_{k-q}-\overline{G}_{k-q}) (F_q-F^\dagger_q+G_q-\overline{G}_q) Q^{xz}_\q Q^{yz}_{\ko+\q}\nonumber\\
&&-i (F_{k+k_0-q}-F^\dagger_{k+k_0-q}+G_{k+k_0-q}-\overline{G}_{k+k_0-q}) (F_q-F^\dagger_q-G_q+\overline{G}_q) Q^{xz}_\q Q^{yz}_{\ko+\q}\nonumber\\
&&-i ({\cal F}_{k+k_0-q}-{\cal F}^\dagger_{k+k_0-q}+{\cal G}_{k+k_0-q}-\overline{\cal G}_{k+k_0-q}) ({\cal F}_{k_0+q}-{\cal F}^\dagger_{k_0+q}+{\cal G}_{k_0+q}-\overline{\cal G}_{k_0+q}) Q^{xz}_{\ko+\q} Q^{yz}_{\ko+\q}\nonumber\\
&&+({\cal F}_{k+k_0-q}+{\cal F}^\dagger_{k+k_0-q}-{\cal G}_{k+k_0-q}-\overline{\cal G}_{k+k_0-q}) (-F_q+F^\dagger_q-G_q+\overline{G}_q) Q^{yz}_{\kk-\q} Q^{yz}_{\ko+\q}\nonumber\\
&&+(-{\cal F}_{k_0+q}+{\cal F}^\dagger_{k_0+q}-{\cal G}_{k_0+q}+\overline{\cal G}_{k_0+q}) (F_{k-q}+F^\dagger_{k-q}-G_{k-q}-\overline{G}_{k-q}) Q^{yz}_{\kk+\ko-\q} Q^{yz}_{\ko+\q}\nonumber\\
&&+({\cal F}_{k_0+q}+{\cal F}^\dagger_{k_0+q}-{\cal G}_{k_0+q}-\overline{\cal G}_{k_0+q}) (-F_{k-q}+F^\dagger_{k-q}-G_{k-q}+\overline{G}_{k-q}) Q^{yz}_\q Q^{yz}_{\ko+\q}\nonumber\\
&&\left.+(-{\cal F}_{k+k_0-q}+{\cal F}^\dagger_{k+k_0-q}-{\cal G}_{k+k_0-q}+\overline{\cal G}_{k+k_0-q}) (F_q+F^\dagger_q-G_q-\overline{G}_q) Q^{yz2}_{\ko+\q}
\right).\nonumber
\end{eqnarray} 

\begin{eqnarray}
&&{\cal P}_\kk-{\cal P}^\dagger_\kk+{\cal S}_\kk-\overline{\cal S}_\kk = \\
&&-\frac{S}{2N} T\sum_{\omega_q,\bf q}
\left(
-2 ({\cal F}_{k-q}+{\cal G}_{k-q}) (F^\dagger_q-\overline{G}_q) Q^{xz}_\kk Q^{xz}_{\kk-\q}-2 (F_{k+k_0-q}+G_{k+k_0-q}) ({\cal F}^\dagger_{k_0+q}-\overline{\cal G}_{k_0+q}) Q^{xz}_\kk Q^{xz}_{\kk+\ko-\q}
\right.\nonumber\\
&&+2 ({\cal F}_{k-q}-{\cal G}_{k-q}) (F^\dagger_q+\overline{G}_q) Q^{xz}_\kk Q^{xz}_\q+({\cal F}_{k-q}+{\cal F}^\dagger_{k-q}+{\cal G}_{k-q}+\overline{\cal G}_{k-q}) (F_q-F^\dagger_q-G_q+\overline{G}_q) Q^{xz}_{\kk-\q} Q^{xz}_\q\nonumber\\
&&+({\cal F}_{k_0+q}-{\cal F}^\dagger_{k_0+q}-{\cal G}_{k_0+q}+\overline{\cal G}_{k_0+q}) (F_{k+k_0-q}+F^\dagger_{k+k_0-q}+G_{k+k_0-q}+\overline{G}_{k+k_0-q}) Q^{xz}_{\kk+\ko-\q} Q^{xz}_\q\nonumber\\
&&+({\cal F}_{k-q}-{\cal F}^\dagger_{k-q}-{\cal G}_{k-q}+\overline{\cal G}_{k-q}) (F_q+F^\dagger_q+G_q+\overline{G}_q) Q^{xz2}_\q+2 (F_{k+k_0-q}-G_{k+k_0-q}) ({\cal F}^\dagger_{k_0+q}+\overline{\cal G}_{k_0+q}) Q^{xz}_\kk Q^{xz}_{\ko+\q}\nonumber\\
&&+({\cal F}_{k_0+q}+{\cal F}^\dagger_{k_0+q}+{\cal G}_{k_0+q}+\overline{\cal G}_{k_0+q}) (F_{k+k_0-q}-F^\dagger_{k+k_0-q}-G_{k+k_0-q}+\overline{G}_{k+k_0-q}) Q^{xz}_\q Q^{xz}_{\ko+\q}\nonumber\\
&&
+4 i ({\cal F}_{k-q} {\cal F}^\dagger_{k_0+q} 
+F_{k+k_0-q} F^\dagger_q 
+{\cal G}_{k-q} \overline{\cal G}_{k_0+q} 
+G_{k+k_0-q} \overline{G}_q) Q^{xz}_\kk Q^{yz}_\kk
\nonumber\\
&&+2 i ({\cal F}^\dagger_{k_0+q}+{\cal G}_{k_0+q}) ({\cal F}_{k-q}+\overline{\cal G}_{k-q}) Q^{xz}_\q Q^{yz}_\kk+2 i ({\cal F}^\dagger_{k-q}+{\cal G}_{k-q}) ({\cal F}_{k_0+q}+\overline{\cal G}_{k_0+q}) Q^{xz}_\q Q^{yz}_\kk\nonumber\\
&&+2 i (F^\dagger_q+G_q) (F_{k+k_0-q}+\overline{G}_{k+k_0-q}) Q^{xz}_\q Q^{yz}_\kk+2 i (F^\dagger_{k+k_0-q}+G_{k+k_0-q}) (F_q+\overline{G}_q) Q^{xz}_\q Q^{yz}_\kk\nonumber\\
&&+2 i (F_{k+k_0-q}-G_{k+k_0-q}) (F^\dagger_q-\overline{G}_q) Q^{xz}_\kk Q^{yz}_{\kk-\q}
+2 i (F_{k+k_0-q}-G_{k+k_0-q}) (F^\dagger_q-\overline{G}_q) Q^{xz}_\kk Q^{yz}_{\ko+\q}\nonumber\\
&&-i (F_{k+k_0-q}-F^\dagger_{k+k_0-q}-G_{k+k_0-q}+\overline{G}_{k+k_0-q}) (F_q-F^\dagger_q-G_q+\overline{G}_q) Q^{xz}_\q Q^{yz}_{\kk-\q}\nonumber\\
&&+2 i ({\cal F}_{k-q}-{\cal G}_{k-q}) ({\cal F}^\dagger_{k_0+q}-\overline{\cal G}_{k_0+q}) Q^{xz}_\kk Q^{yz}_{\kk+\ko-\q}
+2 i ({\cal F}_{k-q}-{\cal G}_{k-q}) ({\cal F}^\dagger_{k_0+q}-\overline{\cal G}_{k_0+q}) Q^{xz}_\kk Q^{yz}_\q\nonumber\\
&&-i ({\cal F}_{k-q}-{\cal F}^\dagger_{k-q}-{\cal G}_{k-q}+\overline{\cal G}_{k-q}) ({\cal F}_{k_0+q}-{\cal F}^\dagger_{k_0+q}-{\cal G}_{k_0+q}+\overline{\cal G}_{k_0+q}) Q^{xz}_\q Q^{yz}_{\kk+\ko-\q}\nonumber\\
&&-i ({\cal F}_{k-q}-{\cal F}^\dagger_{k-q}-{\cal G}_{k-q}+\overline{\cal G}_{k-q}) ({\cal F}_{k_0+q}-{\cal F}^\dagger_{k_0+q}-{\cal G}_{k_0+q}+\overline{\cal G}_{k_0+q}) Q^{xz}_\q Q^{yz}_\q\nonumber\\
&&-i (F_{k-q}+F^\dagger_{k-q}+G_{k-q}+\overline{G}_{k-q}) (F_q+F^\dagger_q-G_q-\overline{G}_q) Q^{xz}_{\kk-\q} Q^{yz}_{\ko+\q}\nonumber\\
&&-i ({\cal F}_{k+k_0-q}+{\cal F}^\dagger_{k+k_0-q}+{\cal G}_{k+k_0-q}+\overline{\cal G}_{k+k_0-q}) ({\cal F}_{k_0+q}+{\cal F}^\dagger_{k_0+q}-{\cal G}_{k_0+q}-\overline{\cal G}_{k_0+q}) Q^{xz}_{\kk+\ko-\q} Q^{yz}_{\ko+\q}\nonumber\\
&&-i (F_{k-q}-F^\dagger_{k-q}-G_{k-q}+\overline{G}_{k-q}) (F_q-F^\dagger_q+G_q-\overline{G}_q) Q^{xz}_\q Q^{yz}_{\ko+\q}\nonumber\\
&&-i (F_{k+k_0-q}-F^\dagger_{k+k_0-q}-G_{k+k_0-q}+\overline{G}_{k+k_0-q}) (F_q-F^\dagger_q-G_q+\overline{G}_q) Q^{xz}_\q Q^{yz}_{\ko+\q}\nonumber\\
&&-i ({\cal F}_{k+k_0-q}-{\cal F}^\dagger_{k+k_0-q}-{\cal G}_{k+k_0-q}+\overline{\cal G}_{k+k_0-q}) ({\cal F}_{k_0+q}-{\cal F}^\dagger_{k_0+q}+{\cal G}_{k_0+q}-\overline{\cal G}_{k_0+q}) Q^{xz}_{\ko+\q} Q^{yz}_{\ko+\q}\nonumber\\
&&-2 (F^\dagger_q-G_q) ({\cal F}_{k+k_0-q}+\overline{\cal G}_{k+k_0-q}) Q^{yz}_\kk Q^{yz}_{\ko+\q}+2 (F^\dagger_{k-q}+G_{k-q}) ({\cal F}_{k_0+q}-\overline{\cal G}_{k_0+q}) Q^{yz}_\kk Q^{yz}_{\ko+\q}\nonumber\\
&&-2 ({\cal F}^\dagger_{k_0+q}-{\cal G}_{k_0+q}) (F_{k-q}+\overline{G}_{k-q}) Q^{yz}_\kk Q^{yz}_{\ko+\q}+2 ({\cal F}^\dagger_{k+k_0-q}+{\cal G}_{k+k_0-q}) (F_q-\overline{G}_q) Q^{yz}_\kk Q^{yz}_{\ko+\q}\nonumber\\
&&+(-{\cal F}_{k+k_0-q}+{\cal F}^\dagger_{k+k_0-q}+{\cal G}_{k+k_0-q}-\overline{\cal G}_{k+k_0-q}) (F_q+F^\dagger_q-G_q-\overline{G}_q) Q^{yz}_{\kk-\q} Q^{yz}_{\ko+\q}\nonumber\\
&&+({\cal F}_{k_0+q}+{\cal F}^\dagger_{k_0+q}-{\cal G}_{k_0+q}-\overline{\cal G}_{k_0+q}) (-F_{k-q}+F^\dagger_{k-q}+G_{k-q}-\overline{G}_{k-q}) Q^{yz}_{\kk+\ko-\q} Q^{yz}_{\ko+\q}\nonumber\\
&&+({\cal F}_{k_0+q}+{\cal F}^\dagger_{k_0+q}-{\cal G}_{k_0+q}-\overline{\cal G}_{k_0+q}) (-F_{k-q}+F^\dagger_{k-q}+G_{k-q}-\overline{G}_{k-q}) Q^{yz}_\q Q^{yz}_{\ko+\q}\nonumber\\
&&\left.+(-{\cal F}_{k+k_0-q}+{\cal F}^\dagger_{k+k_0-q}+{\cal G}_{k+k_0-q}-\overline{\cal G}_{k+k_0-q}) (F_q+F^\dagger_q-G_q-\overline{G}_q) Q^{yz2}_{\ko+\q}
\right).\nonumber
\end{eqnarray} 

\section{Analysis of the decay and confluence processes}
\label{ap_damp}

Omitting the dipolar interaction, the spectrum of Heisenberg AF has the form $\epsilon_\kk = D k-DL_2k^3$ at $k\ll1$ (cf.\ Eq.~\eqref{spectr}). One leads to the following expressions for the decay and confluence processes, respectively, using this spectrum:
\begin{eqnarray}
\label{dec1}
\epsilon_\kk-\epsilon_\q-\epsilon_{\kk-\q} &=&
D(k-q-|\kk-\q|)-3DL_2kq(k-q) = -\frac{D k q_{\perp}^2}{2q(k-q)}-3DL_2kq(k-q)<0,\\
\label{con1}
\epsilon_\kk - \epsilon_\q + \epsilon_{\q-\kk} &=&
D(k-q+|\q-\kk|) + 3DL_2kq(q-k) = \frac{D k q_{\perp}^2}{2q(q-k)} + 3DL_2kq(q-k)>0,
\end{eqnarray}
where $\q_\|$ and $\q_\perp$ are components of $\q$ which are parallel and perpendicular to $\kk$ (see Fig.~\ref{vec}), respectively, and we assume that $\bf q$ and $\bf k$ are nearly parallel each other (i.e., $q_\|\approx q$ and $q_\perp\ll q$). It is seen from Eqs.~\eqref{dec1} and \eqref{con1} that both confluence and decay processes are impossible without dipolar forces. 

\subsection{Decay processes}

Among 8 allowed decay processes \eqref{decay} only the following ones appear to be possible if we take into account the dipolar forces:
\begin{eqnarray}
\label{1}
\epsilon_\kk^+-\epsilon_\q^--\epsilon_{\kk-\q}^-&=&
-3D L_2 k^3 \left(\qd(1-\qd)+\zeta \frac{1-\qd(1-\qd)}{ \qd (1-\qd)}-\eta\right) 
-\frac{D q_{\perp}^2}{2k\qd(1-\qd)}=0,\\
\label{2}
\epsilon_\kk^+-\epsilon_\q^--\epsilon_{\kk-\q}^+&=&
-3D L_2 k^3 \left(\qd(1-\qd)+\zeta \frac{1-\qd(1-\qd)}{ \qd (1-\qd)}-\eta \qd\right)
-\frac{D q_{\perp}^2}{2k\qd(1-\qd)}=0,\\
\label{3}
\epsilon_\kk^+-\epsilon_\q^+-\epsilon_{\kk-\q}^-&=&
-3D L_2 k^3 \left(\qd(1-\qd)+\zeta \frac{1-\qd(1-\qd)}{ \qd (1-\qd)}-\eta(1-\qd)\right)
-\frac{D q_{\perp}^2}{2k\qd(1-\qd)}=0,
\end{eqnarray}
where $\qd = q/k$, $0<\qd<1$, and
\begin{eqnarray}
\zeta &=& \frac{\Delta^2}{6L_2D^2k^4}=\frac{\Delta^2}{72L_2J^2S^2k^4},\\
\label{eta}
\eta &=& \frac{\omega_0 L_1(\varphi_\kk)\sin^2\theta_\kk}{72L_2J k^2}.
\end{eqnarray}
It is seen that Eq.~\eqref{1} can have a solution if the following inequality holds:
\begin{equation}
\label{eq}
\frac{1}{z}\left(z^2+\zeta(1-z)-\eta z  \right)<0,
\end{equation} 
where $z=\qd(1-\qd)$ and $0<z<1/4$. Solving the quadratic equation, one finds that Eq.~\eqref{eq} is satisfied when
\begin{eqnarray}
\label{z1}
z_- &<& \qd(1-\qd)<z_+,\\
\label{z12}
z_\pm &=& \frac12\left(\zeta+\eta\pm\sqrt{\left(\zeta+\eta\right)^2-4\zeta}\right).
\end{eqnarray}
It is convenient to discuss a limiting case of $\eta\gg \zeta$ that reads
\begin{equation}
k\gg \sqrt{\frac{J}{\omega_0}}\frac{\Delta}{D\sin\theta_\kk}.
\label{kth_cond}
\end{equation}
The opposite limit of $\eta\ll \zeta$ has no meaning because it could be realized for $\zeta>4$ only in which case $z>1$. One has from Eq.~\eqref{z12} at $\eta\gg \zeta$ 
\begin{equation}
z_+\approx \eta, \quad z_-\approx \frac{\zeta}{\eta}.
\label{z2}
\end{equation}
The requirement $z_+\gg z_-$ reads
\begin{equation}
\sin\theta_\kk\gg\sqrt{\frac{\Delta}{S\omega_0}}.
\end{equation}
It is seen from Eq.~\eqref{eta} that $\eta\ll1$ if \eqref{kth_cond} holds. As a result there are two intervals for $\qd$ inside which inequality \eqref{z1} is satisfied:
\begin{eqnarray}
\label{int1}
\qd &\in& \left(\frac{\zeta}{\eta},\eta\right),\\
\label{int2}
\qd &\in& \left(1-\eta,1-\frac{\zeta}{\eta}\right).
\end{eqnarray}

It is easy to demonstrate that solutions of Eqs.~\eqref{2} and \eqref{3} exist only for $\qd$ lying inside intervals \eqref{int2} and \eqref{int1}, respectively.

\subsection{Confluence processes}

Possible confluence processes have the form
\begin{eqnarray}
\label{c1}
\epsilon_\kk^-+\epsilon_{\q-\kk}^+-\epsilon_\q^+&=&
3D L_2 k^3 \left(\qd(\qd-1)+\zeta \frac{\qd(\qd-1)+1}{\qd (\qd-1)}-\eta\right)
+\frac{D q_{\perp}^2}{2k\qd(\qd-1)}=0,\\
\label{c2}
\epsilon_\kk^-+\epsilon_{\q-\kk}^--\epsilon_\q^+&=&
3D L_2 k^3 \left(\qd(\qd-1)+\zeta \frac{\qd(\qd-1)+1}{\qd (\qd-1)}-\eta \qd\right)
+\frac{D q_{\perp}^2}{2k\qd(\qd-1)}=0,\\
\label{c3}
\epsilon_\kk^++\epsilon_{\q}^--\epsilon_{\q-\kk}^+&=&
3D L_2 k^3 \left(\qd(\qd-1)+\zeta \frac{\qd(\qd-1)+1}{\qd (\qd-1)}-\eta\qd\right)
+\frac{D q_{\perp}^2}{2k\qd(\qd-1)}=0,
\end{eqnarray}
and there are also other three processes which differ from the presented ones by replacement of $\q$ by $\kk-\q$. Comparing Eqs.~\eqref{1}--\eqref{3} and \eqref{c1}--\eqref{c3} one concludes that it is necessary to analyze similar inequality on $z=\qd (\qd-1)$
\begin{equation}
\label{eqc}
\frac{1}{z}\left(z^2+\zeta(1+z)-\eta z  \right)<0.
\end{equation} 
Inequality \eqref{eqc} is satisfied if
\begin{eqnarray}
\label{z1c}
z_- &<& \qd(\qd-1)<z_+,\\
\label{z12c}
z_\pm &=& \frac12\left(\eta-\zeta\pm\sqrt{\left(\eta-\zeta\right)^2-4\zeta}\right).
\end{eqnarray}
One has in the limiting case of $\eta\gg \zeta$ 
\begin{equation}
\label{int1c}
\qd \in \left(1+\frac{\zeta}{\eta},1+\eta\right)
\end{equation}
so that $q\sim k$ inside this interval.

The opposite limiting case of $\eta\ll \zeta$ is also possible for Eqs.~\eqref{c2} and \eqref{c3}. This limiting case corresponds to $q\gg k$ and
\begin{equation}
\frac\Delta D\ll k\ll \sqrt{\frac{J}{\omega_0}}\frac{\Delta}{D\sin\theta_\kk}.
\end{equation}
In order Eqs.~\eqref{c2} and \eqref{c3} have solutions, $\qd$ should lie between roots of the equation $z^2-\eta z+\zeta=0$, i.e., in the interval
\begin{equation}
\qd \in \left( \frac\eta2\left(1-\sqrt{1-\frac{4\zeta}{\eta^2}}\right), \frac\eta2\left(1+\sqrt{1-\frac{4\zeta}{\eta^2}}\right) \right).
\end{equation}
Quantities $A$ and $B$ defined by Eqs.~\eqref{a} and \eqref{b}, respectively, are related to $\zeta$ and $\eta$ as follows: $A=\eta/2$ and $B=4\zeta/\eta^2$.

\section{Competing single-ion anisotropy and effect of impurities}
\label{defect}

We discuss in this appendix the effect of a cubic magnetocrystalline anisotropy on the properties of Heisenberg AF with dipolar forces considered in the main text. Assuming for simplicity that $S$ is large, one can model the effect of the cubic anisotropy by the following single-ion interaction:
\begin{equation}
\label{anis}
	\frac{A}{S^2}\sum_i \left((S_i^x)^2(S_i^y)^2 + (S_i^x)^2(S_i^z)^2+(S_i^y)^2(S_i^z)^2\right).
\end{equation}
We imply below that $\omega_0^2/SJ\sim|A|\ll\omega_0\ll J$. Let us assume also that the anisotropy constant $A'$ and the spin value $S'$ differ from $A$ and $S$, respectively, at some randomly distributed sites, which concentration is equal to $x\ll1$. As a result, bilinear part of the Hamiltonian \eqref{h2} acquires the following correction if the staggered magnetization is directed along a cube edge:
\begin{equation}
\label{h2p}
\mathcal{H}_2^{anis} =
2AS\sum\limits_\kk a^\dagger_\kk a_\kk
+ 2(\alpha A'-A)S\sum\limits_n \chi(n) a^\dagger_n a_n,
\end{equation}
where $\alpha=S'/S$, $\chi(n)=1$ at sites occupied by impurities, and $\chi(n)=0$ at other sites. Discussion of the Hamiltonian $\mathcal{H}_2+\mathcal{H}_2^{anis}$ can be carried out in the first order in $x$ using the $T$-matrix approach as it is done, e.g., in Ref.~\cite{wan}. The situation here is simplified greatly by two circumstances: i) $|A|,|A'|\ll J$, and ii) sums over $\kk$ of Green's functions $F(\omega,\kk)$, $F^\dagger(\omega,\kk)$, ${\cal G}(\omega,\kk)$, $\overline{\cal G}(\omega,\kk)$, ${\cal F}(\omega,\kk)$, and ${\cal F}^\dagger(\omega,\kk)$ at $\omega\ll SJ$ are of the order of $\omega_0/SJ^2$ whereas such sum for $G(\omega,\kk)={\overline G}(-\omega,-\kk)^*$ is much greater being of the order of $1/SJ$. As a consequence, the greatest contributions from Eq.~\eqref{h2p} arises in $\Sigma_\kk$ and $\overline{\Sigma}_\kk$. Then, one obtains from Eqs.~\eqref{omega1} and \eqref{h2p} for the correction to $\Omega(\omega,\kk)$ at $k\ll1$:
\begin{equation}
\label{eren}
	\mp\frac{\Omega_{anis}\left(\omega=\epsilon_{0\kk}^\pm+i\delta,\kk\right)}{2\sqrt{d_\kk}} 
	=
	4S^2J_{\bf 0} 
	\left( A + x(\alpha A'-A) 
	+ 
	2x(\alpha A'-A)^2S\frac1N\sum_{\bf q} \frac{SJ_{\bf 0}}{(\epsilon_{0\kk}^\pm+i\delta)^2-S^2(J_{\bf 0}^2-J_{\bf q}^2)}
	\right).
\end{equation}
To derive the last term in Eq.~\eqref{eren}, we set $\omega_0=0$ in $\sum_\q (G(\omega,\q)+{\overline G}(\omega,\q))$ and use Eq.~\eqref{gc} for the normal Green's function. The imaginary part of the last term in Eq.~\eqref{eren} determines the magnon damping due to the scattering on impurities whereas its real part is negligibly small compared to the second term because $|A|,|A'|\ll J$. Using Eqs.~\eqref{specsq} and \eqref{eren}, we obtain for the square of the gap in the spectrum
\begin{equation}
\label{specimp}
\tilde\Delta^2 = \Delta^2 + 4S^2J_{\bf 0} \left( A + x(\alpha A'-A) \right),
\end{equation}
where $\Delta$ is the contribution to the gap from dipolar forces given by Eq.~\eqref{gap}. The magnon damping due to the scattering on impurities is estimated from Eqs.~\eqref{dampdef} and \eqref{eren} as
\begin{equation}
\label{gamimp}
\Gamma_\kk \sim x S\frac{(\alpha A'-A)^2}{J}.
\end{equation}

It is seen from Eq.~\eqref{specimp} that the gap in the spectrum can vanish if the anisotropies of dipolar origin \eqref{de} (accompanied by the gap $\Delta$), $A$ and $A'$ compete. The change of the sign of $\tilde\Delta^2$ signifies that the easy direction switches from a cube edge to a cube space diagonal. For instance, the competition arises in ${\rm TlMn}_{1-x}{\rm Co}_x{\rm F}_3$ and ${\rm RbMn}_{1-x}{\rm Co}_x{\rm F}_3$ between $A<0$ (favoring the cube space diagonals) on the one hand and $A'>0$ and the dipolar anisotropy (favoring the cube edges) on the other hand. The gap \eqref{specimp} vanishes in this case when $x$ is equal to
\begin{equation}
\label{xs}
x_*=\frac{\Delta^2 + 4S^2J_{\bf 0} A}{4S^2J_{\bf 0}(A-\alpha A')}.
\end{equation}
Notice that $x_*$ given by Eq.~\eqref{xs} is much smaller than unity in ${\rm TlMn}_{1-x}{\rm Co}_x{\rm F}_3$ and ${\rm RbMn}_{1-x}{\rm Co}_x{\rm F}_3$ because $A'\gg |A|$ and $\Delta^2\sim 4S^2J_{\bf 0} |A|$.

The above consideration is justified when $\epsilon_\kk\gg\Gamma_\kk$. It is seen from Eq.~\eqref{gamimp} that this condition can be invalid for some $\kk$. For instance, $\epsilon_\kk\gg\Gamma_\kk$ at $x\sim x_*$ if
\begin{equation}
\label{impin}
	k\gg x\frac{(\alpha A'-A)^2}{J^2}.
\end{equation}
The invalidity of the inequality $\epsilon_\kk\gg\Gamma_\kk$ can signify a localization of states near the magnon band bottom (see, e.g., Ref.~\cite{utesov} and references therein). Then, Eq.~\eqref{xs} is just an estimation of the concentration value at which the easy direction switches because Eq.~\eqref{specimp} is invalid when $\tilde\Delta$ is smaller than $\Gamma_\kk$ given by Eq.~\eqref{gamimp}. In ${\rm TlMn}_{1-x}{\rm Co}_x{\rm F}_3$ and ${\rm RbMn}_{1-x}{\rm Co}_x{\rm F}_3$ at $x\sim x_*$, inequality \eqref{impin} reads as $k\gg 10^{-6}$~\AA$^{-1}$. As soon as the great damping due to magnon interaction is expected in these compounds at $k\lesssim 10^{-3}\div10^{-4}$~\AA$^{-1}$, these materials can be suitable for the experimental observation of the magnon breakdown discussed in the main text.

It should be noted also that as soon as defects change the bare spectrum considerably at $x\approx x_*$ and $k\ll1$, the results obtained above must be used with caution. In particular, one cannot exclude the possibility of great spectrum change by terms of higher orders in $x$. It is difficult to analyze the whole series in $x$ but we notice that all the expected contributions are small as compared to those considered above due to the smallness of ratios $A/J$ and $A'/J$.

\bibliography{lit}

\end{document}